\documentclass[useAMS,usenatbib]{mn2e}
\usepackage{graphicx}
\usepackage{epstopdf}
\usepackage{rotating}
\usepackage{caption}
\usepackage{color}
\usepackage{booktabs}

\makeatletter

\newcommand\ion[2]{#1$\;${\small\rmfamily\@Roman{#2}}\relax}

\title[Chemical abundances in the GCs NGC 5024 and NGC 5466 from optical and IR spectroscopy]{Chemical abundances in the globular clusters 
NGC 5024 and NGC 5466 from optical\thanks{Based on observations obtained with the Hobby-Eberly Telescope, which is a joint project of the University of Texas at Austin, the Pennsylvania State University, Stanford University, Ludwig-Maximilians-UniversitŠt MŸnchen, and Georg-August-UniversitŠt Gšttingen.} and infrared spectroscopy}
\author[M.P. Lamb et al.]{M.P. Lamb$^{1,5}$\thanks{E-mail: masen@uvic.ca}, K.A. Venn$^{1}$, M.D. Shetrone$^{2}$, C.M Sakari$^{1,3}$, and B.J. Pritzl$^{4}$\\
$^{1}$Department of Physics and Astronomy, University of Victoria, Victoria, British Columbia, V8W 3P2, Canada\\
$^{2}$Mcdonald Observatory, University of Texas at Austin, HC75 Box 1337-MCD, Fort Davis, TX 79734, USA\\
$^{3}$Department of Astronomy, University of Washington Seattle, Seattle, WA 98195, USA\\
$^{4}$University of Wisconsin Oshkosh, Oshkosh, WI 54901, USA\\
$^{5}$NRC Herzberg Institute of Astrophysics, 5071 West Saanich Road, Victoria, BC V9E 2E7, Canada}

\begin{document}

\pagerange{\pageref{firstpage}--\pageref{lastpage}} \pubyear{2013}

\maketitle

\label{firstpage}

\begin{abstract}
Detailed chemical abundances for five stars in two Galactic globular clusters, NGC 5466 and NGC 5024, are presented from high resolution optical (from the Hobby-Eberley Telescope) 
and infrared spectra (from the SDSS-III APOGEE survey).
We find [Fe/H] = -1.97~$\pm$~0.13 dex for NGC 5466, and [Fe/H] = -2.06~$\pm$~0.13 dex for NGC 5024, and the typical abundance pattern for globular clusters for the remaining elements,
e.g., both show evidence for mixing in their light element abundance ratios (C, N), and AGB contributions in their heavy element abundances (Y, Ba, and Eu).   These clusters were selected to examine chemical trends that may correlate them with the Sgr dwarf galaxy remnant, but at these low \textcolor{black}{metallicities no} obvious differences from the Galactic abundance pattern are found.   Regardless, we compare our results from the optical and infrared analyses to find that oxygen and silicon abundances \textcolor{black}{determined} from the infrared spectral lines are in better agreement with the other alpha-element ratios and with smaller random errors.   

\end{abstract}


\section{Introduction}

The discovery of the accretion of globular clusters from the Sagittarius dwarf galaxy \citep{b85}, has \textcolor{black}{led} to the question as to how many globular clusters have been captured by the Milky Way. Multiple studies have looked at the globular cluster systems of the Milky Way to derive an age-metallicity relationship and have come to different conclusions as to which clusters have likely been accreted \citep{b32,b33,b34,b82}. The question is still open as to which type of clusters are accreted and which form in situ; and furthermore what the fraction of each type is within the Milky Way.

Globular clusters (GC) formed in dwarf galaxies may differ from those found in the Galactic halo, depending on their age and metallicity.   Dwarf galaxies show a wide variety of star formation histories \citep{b35,b36,b37} that are predicted to lead to variations in their metallicity distribution functions and chemical abundances. It has also been suggested these variations could be attributed to differences in the IMFs of these galaxies \citep{b73}. If the IMFs are the root cause of these differences then this would also lead to differences in the age-metallicity relationship, which is observed by both \cite{b32} and \cite{b34}. From observations, field stars in dwarf galaxies do exhibit different abundance ratios from Milky Way (MW) field stars, e.g., lower [$\alpha$/Fe] ratios and variations in neutron capture element ratios at intermediate metallicities\textcolor{black}{. However} these typically do not show up until [Fe/H] $\sim -1.5$ \citep{b86,b87,b13,b27,b41,b42,b28}.   At metallicities below [Fe/H] $= -1.5$ the abundance variations between field and GC stars become less pronounced in dwarfs and the MW \citep{b68, b81,b83,b45}; a good example of this is M54, located at the heart of the Sagittarius (Sgr) dwarf accretion remnant. M54 has a much lower metallicity than the Sgr field stars (e.g., \citealt{b83}) and the [$\alpha$/Fe] ratios resemble the field stars in the MW halo and its detailed chemical abundance ratios resemble the patterns seen in other globular cluster systems (e.g., the Na-O anti-correlation; \citealt{carretta2009}).   Therefore, other than its physical association with the Sgr remnant, M54 does not stand out from other GCs in terms of its chemical abundance patterns, similar to the metal-poor GCs Terzan 8 and Arp 2 (both also kinematically and spatial associated with the Sgr stream \citealt{b88}). On the other hand, Hodge 11 in the LMC at [Fe/H] = -2.0 does have lower [$\alpha$/Fe] than MW field and GC stars\citep{b90}; and Ruprecht 106 has an anomalously low [$\alpha$/Fe] ratio for a MW GC \citep{b48}.

Two metal-poor clusters that have been associated with the Sgr stream are NGC\,5024 (M53) and NGC\,5466 \citep{b50,b51}.    Both of these clusters are more metal-poor than M54 (each at [Fe/H] $\sim -2$, Harris 1996), which means that detailed chemistry could be inconclusive as to their origins in the Sgr dwarf galaxy.   We have opted to study the chemistry in these GCs regardless though because (1) there are few published chemistries for these clusters and (2) they are both associated with other interesting dynamical structures. NGC\,5466 has a large tidal tail \citep{b55}\textcolor{black}{. However} it appears to have no association with the Sgr stream, and knowing the chemistry of this GC can help identify members in the tidal feature. It is also worth nothing that NGC\,5466 has a retrograde orbit, suggestive of an extragalactic origin \citep{b56,b32}. NGC\,5024 may be linked by a stellar bridge to NGC\,5053 \citep{b53,b50}, although no bridge was seen by \cite{b54}, and detailed chemical abundances of stars in these two \textcolor{black}{clusters} can be used to study if their formation was coeval. 

Detailed chemical abundances have been determined for a few stars in these clusters; one star in NGC\,5024 and two stars in NGC\,5466 by \cite{b57} (at lower resolution and S/N than discussed in this paper, see Section \ref{sec:obs}), and one anomalous cepheid in NGC\,5466 by \cite{b30}. Iron abundances for several stars in NGC\,5024 have also been estimated from photometry by \cite{b10}. All of these analyses confirm the metallicities of [Fe/H] $\sim -2$ dex \citep{b7}. Carbon abundances have been derived from CN and CH band strengths for over a dozen stars in both NGC 5024 and NGC\,5466 from \cite{b46} and \cite{b21}, respectively.   In both clusters, large variations in the [C/Fe] ratios are found, typical of stars that have undergone deep mixing on the red giant branch.

In this paper, we determine the chemical composition of individual stars in each GC. Abundances are determined from both optical and infrared spectroscopic data. This larger wavelength coverage allows us to determine the abundances of more spectral lines and more elements (e.g., CN, CO, OH, Si, and Al are ubiquitous in the IR yet rare in the optical). We also compare the accuracy of the abundance results between the two wavelength regions (similar to \citealt{b58}).

\section{Observations and Data Reduction}

\subsection{Observing Program}
\label{sec:obs}
\begin{figure*}
\centering
   \includegraphics[clip=true,trim = 5 5 0 5,width=0.5\textwidth]{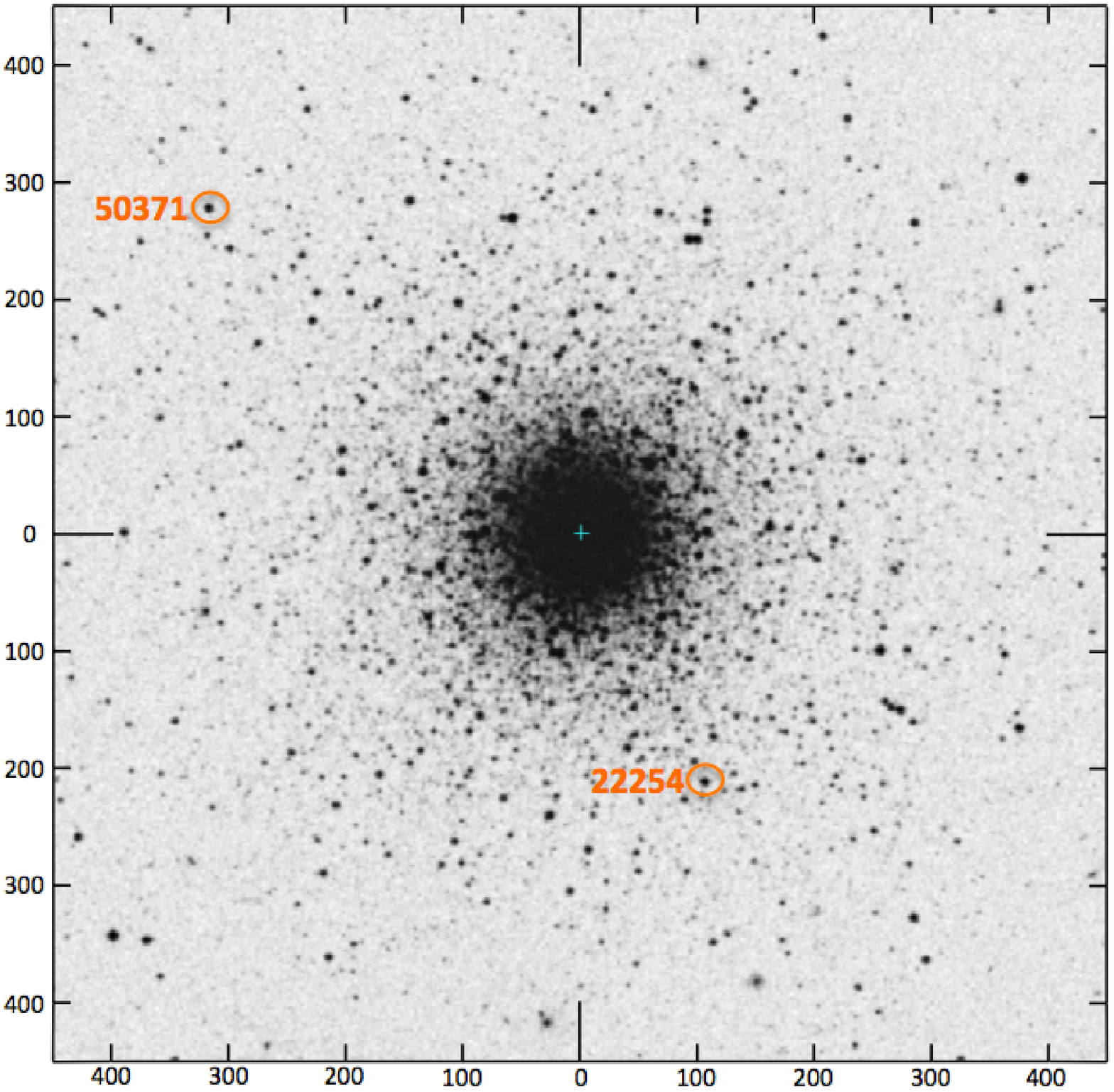}\includegraphics[clip=true,trim = 5 5 0 5,width=0.5\textwidth]{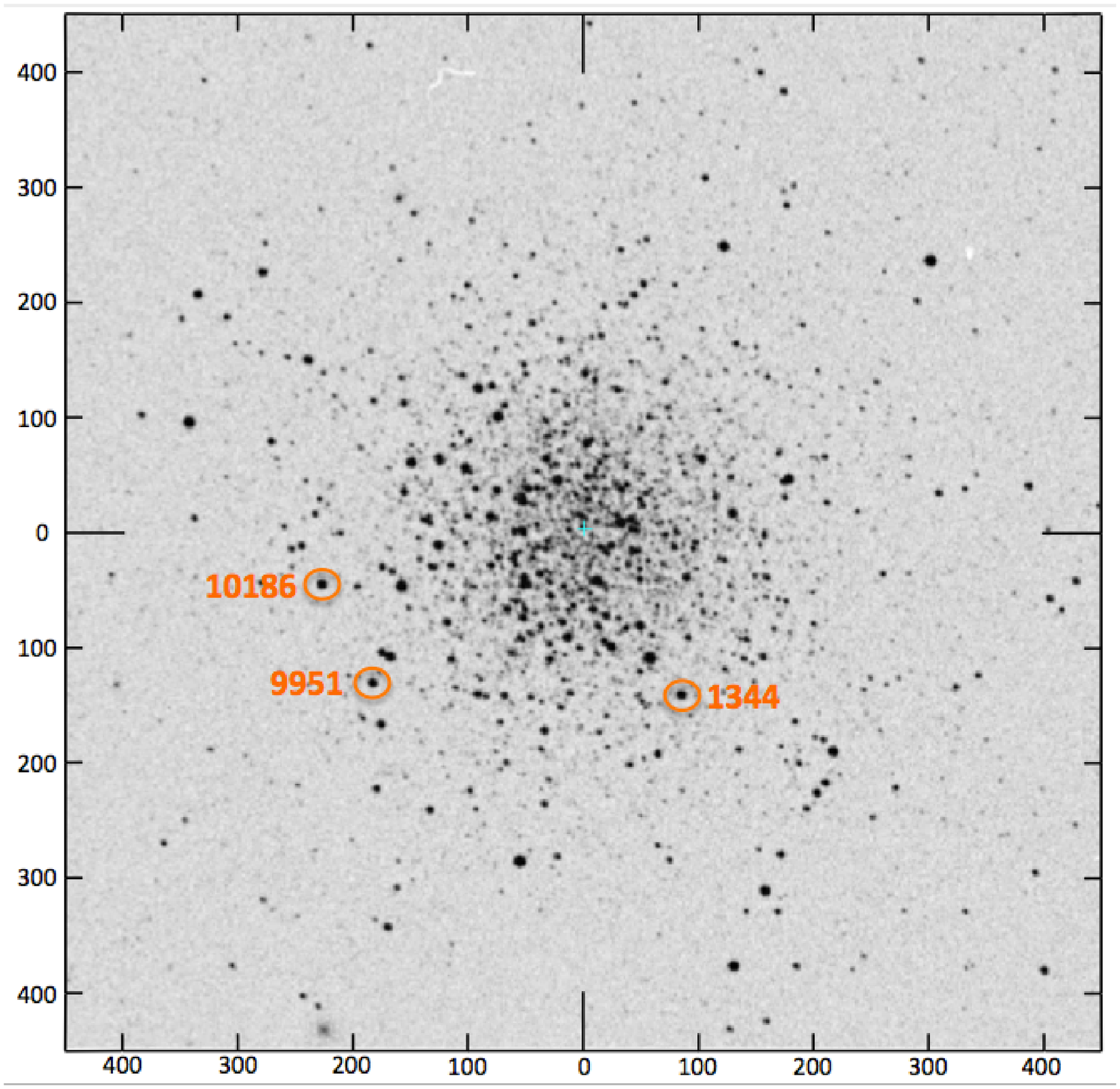}
   \caption{Positions of our science targets in NGC 5024 (left) and NGC 5466 (right). Axes are in arc seconds from the cluster centre (the centre is noted by the cyan cross in each image). North is up and East is left. Images taken from the SDSS survey.}
   \label{fig:finderchart}
\end{figure*}

Five red giant branch (RGB) stars have been selected in the outer regions of two globular clusters, NGC 5024 and NGC 5466, for detailed spectral analyses.     The locations of these objects are shown in Fig. \ref{fig:finderchart} and their fundamental properties are listed in Table~1.   Targets were chosen based on their V magnitudes, and V-I colours from the \textcolor{black}{Stetson database \citep{b9}}.   Foreground contamination is minimal.   

Optical spectra were gathered with the High Resolution Spectrograph (HRS, \citealt{b61}) on the HET\footnote{Observing time was allocated through NOAO program number 05 A-330, via the predecessor of the TSIP program, i.e., the NSF Facilities Instrumentation Program}. The HRS was \textcolor{black}{configured} at resolution R = 30,000 with 2x2 pixel binning using the 2 arcsecond fibre.   The HRS splits the incoming beam onto two CCD chips, from which the spectral regions regions 6000 - 7000 \AA\hspace{0.020 mm}  (red chip) and 4800 - 5900 \AA\hspace{0.020 mm}  (blue chip) were extracted for this work. 
Two standard stars were also observed, RGB stars with previously published spectral analyses in each of the globular clusters M3 and M13. 
The signal to noise (S/N) for these seven targets ranged from 40 - \textcolor{black}{85} (see Table \ref{table:obs}). 
 
IR spectra for four of the five targets in NGC 5466 and NGC 5024 are available in the APOGEE DR10 release\footnote{https://www.sdss3.org/dr10/}. APOGEE \textcolor{black}{provides} H-band spectra, ranging from 15000-17000\AA\hspace{0.020 mm} at a resolution R $\sim$ 20,000 with S/N $\ge$ 100. These spectra expand \textcolor{black}{the} analysis of our main science targets to features at longer wavelengths.    We have not analysed the APOGEE spectra of our M3 and M13 standard stars since there are no published results of their infrared spectral features for a comparison.

\begin{table*}
 \centering
\hskip-3.0cm\begin{minipage}{140mm}
  \caption{The sample of stars observed in the optical with the HET}
  \begin{tabular}{@{}lccrcccr@{}}
  \hline
                                     & 				& 			& 			& 		& Red S/N & Blue S/N & \\
   Star  & R.A. (J2000) & Dec. (J2000) & Date Obs. & Exp. Time (s)\footnote{Total exposure time is listed, including the number of nights that observations were taken.} & @6250\AA\hspace{0.020 mm}  & @5500\AA\hspace{0.020 mm} & $V_{hel.}$ (km/s)\\
    \hline
NGC 5024-22254 	&13 12 47.94		&+18 06 32.1	&May 2005	& 2700(1)	& 50	& 40	& -57.57 $\pm$ 0.50	\\
NGC 5024-50371 	&13 13 17.36 		&+18 14 46.4	&March 2007	& 3387(1)	& 75	& 55 & -62.69 $\pm$ 0.24	\\
NGC 5466-9951 	&14 05 41.08 		&+28 29 48.2	&June 2005	& 6350(2)	& 60 & 55 & 126.53 $\pm$ 1.10\\ 
NGC 5466-1344 	&14 05 20.73 		&+28 29 42.0	&June 2006	& 6600(3)	& 70 & 60 & 129.29 $\pm$ 0.81\\
NGC 5466-10186  	&14 05 44.53 		&+28 31 13.5	&July 2006	& 3300(1)	& 70 & 60 & 107.33 $\pm$ 0.48\\
M3-C41303-2217  	&13 41 30.30 		&+28 29 42.0	&March 2008	& 3000(1)	& 50 & 40 & -145.25 $\pm$ 0.37\\
M13-III-18  		&16 41 24.64 		&+36 25 45.0	&March 2008	& 3000(1)	& 85 & 75 & -233.64 $\pm$ 0.23 \\
\hline
\label{table:obs}
\end{tabular}
\end{minipage}
\end{table*}

\subsection{Optical Data Reduction}

The HET-HRS data were reduced using standard IRAF\footnote{IRAF (Image Reduction and Analysis Facility) is distributed by the National Optical Astronomy Observatory, which is operated by the Association of Universities for Research in Astronomy, Inc., under cooperative agreement with the National Science Foundation.} packages.
Some data was taken over multiple nights\textcolor{black}{. Therefore} data was reduced per night and coadded in those circumstances (see Table 1).   

For each science observation, bias images (5), flat field images (5), a wavelength calibration, and a telluric standard star (a rapidly-rotating hot star) were also obtained.    A medianed bias was subtracted from the red CCD images, but not the blue CCD images at the suggestion of the HET data reduction manuals\footnote{http://hydra.as.utexas.edu/?a=help\&h=29\#HRS} as this CCD is very clean and bias subtracting can add noise.    Three science exposures from 2005 were taken with an older red CCD that suffered from hot pixels and bad columns - these exposures were corrected in IRAF (using \textit{fixpix}).   Images were divided by a medianed flat field, and scattered light was removed before aperture extraction.   Spectra were extracted with variance weighting (to reduce cosmic ray contamination),
and wavelength calibrated using a standard thorium-argon lamp.    The telluric standards were reduced by the same methods, and divided into the science spectra to 
remove atmospheric features.    The sky fibre on the HRS was not used because our science targets are quite bright.   

\begin{figure}
   \centering
    \includegraphics[width=0.5\textwidth]{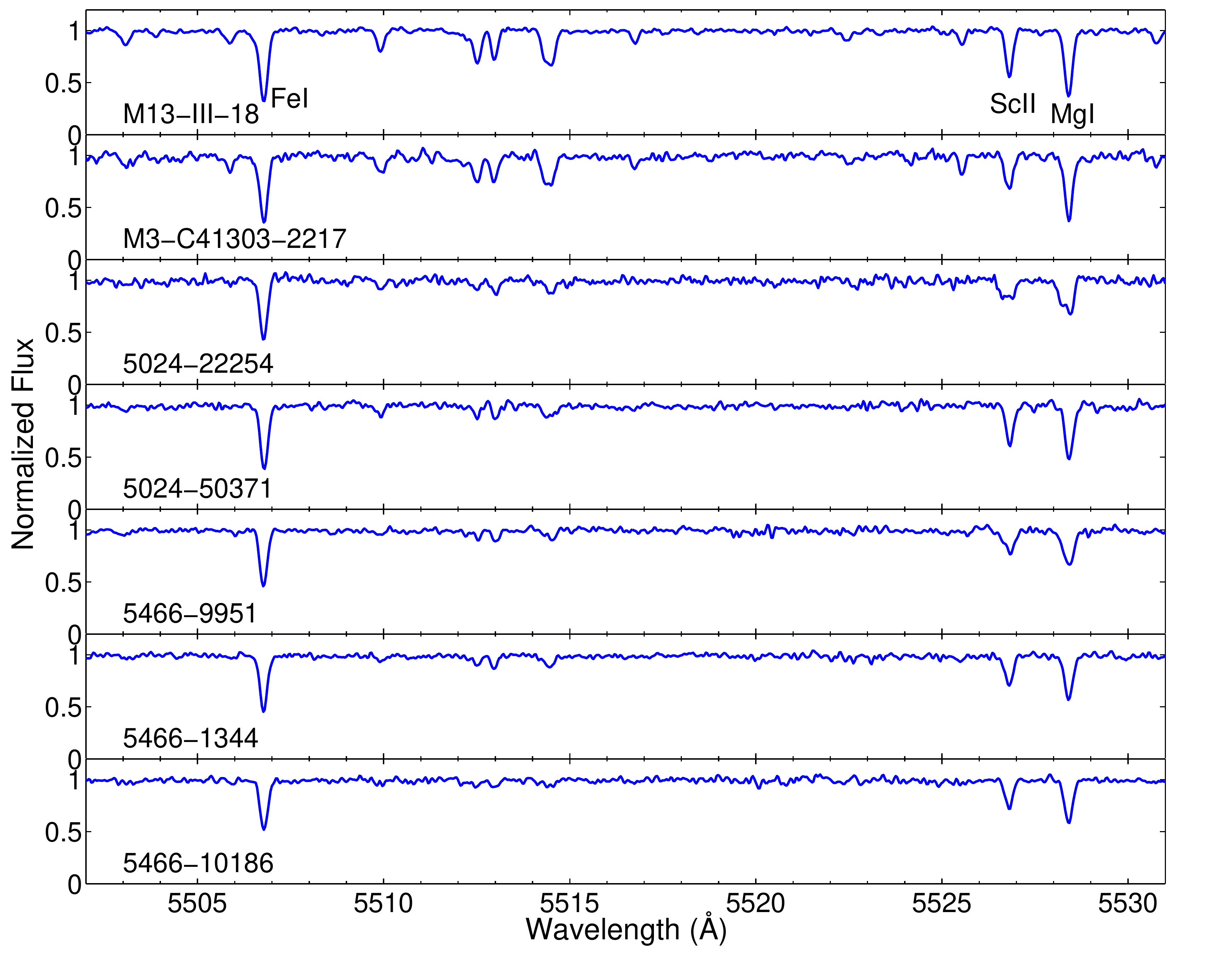}
    \includegraphics[width=0.5\textwidth]{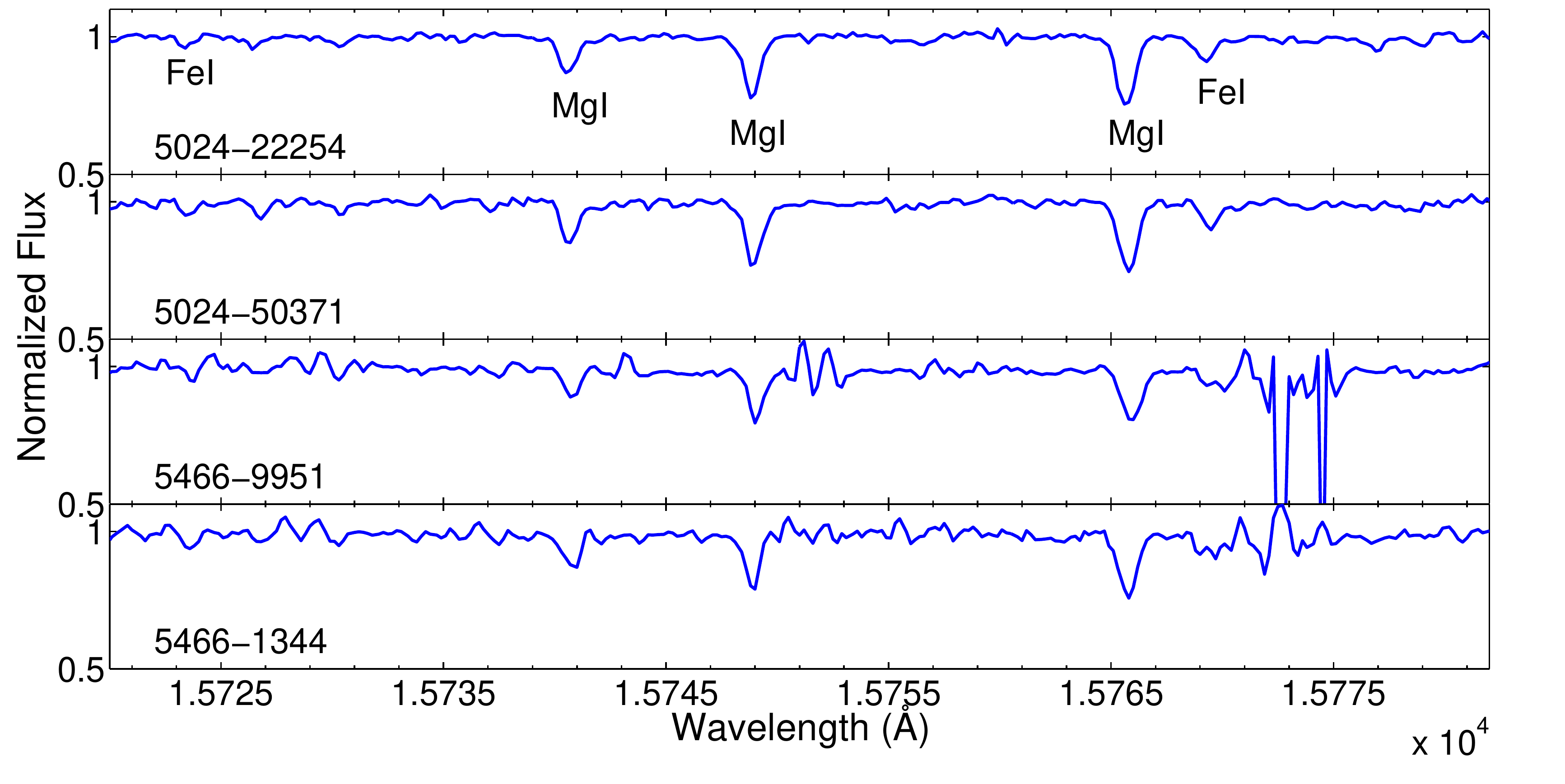}
   \caption{Sample spectral regions in the optical (blue chip, top), and infrared (bottom) showing magnesium, scandium, and iron lines that were used in the abundance analysis.}
   \label{fig:mgspectra}
\end{figure}

Samples of the final spectra are shown in Fig. \ref{fig:mgspectra}.  Radial velocities were calculated using the IRAF task \textit{fxcor}, and cross-correlating the heliocentric-corrected spectrum of Arcturus \citep{b26}.   These values are listed in Table \ref{table:obs}. 
The velocities in NGC 5466 are in excellent agreement with those calculated per star by \cite{b21}\textcolor{black}{. Thus} this cluster shows a significant dispersion (Shetrone et al. 2012 finds a velocity dispersion of $\sim$17 km/s from 67 stars). 

Post-pipeline processed APOGEE spectra are available in the SDSS DR10 database, where they have been reduced and coadded when multiple exposures were taken, and then continuum normalized.   We took and additional step and shifted these spectra from the vacuum rest frame to the air rest frame. 

\section{Equivalent Width Analysis of Optical Spectra}

Optical abundances are determined from an equivalent width (EW) analysis. All equivalent widths were initially measured using DAOSPEC \citep{b1}\footnote{DAOSPEC is a program that is capable of measuring equivalent widths and radial velocities of spectra. It was written by P.B. Stetson for the Dominion Astrophysical Observatory, National Research Council, Canada.}. The atomic lines used with DAOSPEC are comprised of a list taken from several sources: \citet{b13}, \citet{b14}, \citet{b15}, \citet{b16}, \citet{b17}, \citet{b18}, \citet{b19}, and \citet{b77}. \textcolor{black}{Discrepancies between DAOSPEC and splot occur for lines with EWs greater than 100 m\AA\hspace{0.20 mm}}; the IRAF measurements were adopted for lines $>$ 100 m\AA. Lines were also examined by eye for unrecognized blends or other mismeasurements (i.e. contaminated with noise spikes, etc.), using the spectrum of Arcturus and the Sun as references. Lines with EWs $>$ 200 m\AA\hspace{0.20 mm} were not analysed since they tend to be very sensitive to small uncertainties in \textcolor{black}{microturbulence and other stellar atmosphere effects.}

\subsection{$\Delta EW$}

The uncertainty in EW measurements can be estimated from the revised Cayrel formula \citep{b78,b11},

\begin{equation}
\Delta (EW) \sim \frac{\sqrt{1.5 \cdot FWHM \cdot \Delta x}}{S/N}
 \end{equation}
 where FWHM is the measured full width at half maximum of a particular line in \AA, $\Delta$ x is the dispersion of the spectra in \AA /pixel (37.9  m\AA\hspace{0.20 mm} for the Blue chip and 50.89 m\AA\hspace{0.20 mm} for the Red), and S/N is the signal-to-noise measured in the wavelength region of the particular line being measured. For a stronger line (EW $\sim 100$ m\AA\hspace{0.20 mm}) then \textcolor{black}{$\Delta$(EW) = 2.7 m\AA\hspace{0.20 mm} ($\sim$ 3\% error)} whereas weaker lines (EW $\sim$ 20 m\AA\hspace{0.20 mm}) yield a value closer to 10\% \textcolor{black}{error}\textcolor{black}{. Thus} we conservatively adopt 10\% as the uncertainty in EW measurements for each line.

\begin{table*}
\centering
\begin{minipage}{140mm}
\caption{Equivalent Widths and Atomic Data}
\begin{tabular}{@{}lccrccccccr@{}}
\hline

	&  			& 		& 		&\multicolumn{2}{c}{NGC 5024}&  \multicolumn{3}{c}{NGC 5466}						& \multicolumn{1}{c}{M3}&  \multicolumn{1}{c}{M13}	\\ \cmidrule(r){5-6} \cmidrule(r){7-9}  \cmidrule(r){10-10}  \cmidrule(r){11-11}
	& $\lambda$ 	& $\chi$ 	& 			& 22254			& 50371 			& 1344			& 	9951		& 10186			& C41303-2217		&  III-18  \\
   Ion	& (\AA)		& (eV)	&$\log$ $gf$ 	& (m\AA)			& (m\AA)			& (m\AA)			& (m\AA)		& (m\AA)			& (m\AA)		& (m\AA)\\
\hline
Fe I	& 5302.302	& 3.28 	& -0.880		& 78.4			& 94.9			&83.4			& 	 78.5		& 75.5			& 109.0		&  123.7\\
Fe I	& 5307.370      	& 1.61    	& -2.912		& 68.8			& 93.3			& 91.1			& 	83.7		& 74.5			& 123.1		&  122.2\\
Fe I	& 5324.190     	& 3.21    	& -0.100		& 123.3			& 123.9			&111.8			& 	 123.8	& 111.5			& 158.4		&  158.2\\
Fe I	& 5339.930	& 3.27 	& -0.720		& 95.9			& 92.7			&79.3			& 	 75.3		& 77.1			& 123.8		&  127.3\\
Fe I	& 5364.860	& 4.45 	&  0.228		& 54.4			& 56.8			&57.3			& 	 61.7		& 54.3			& 86.8		&    93.3\\
...\\
\hline

\label{table:EWs}
\end{tabular}
*Full table available online
\end{minipage}
\end{table*}

\subsection{EW Comparison with Standard Stars}
\label{sec:EWcomp}
\begin{figure}
   \centering
   \includegraphics[clip=true,trim = 5 0 0 0,width=0.48\textwidth]{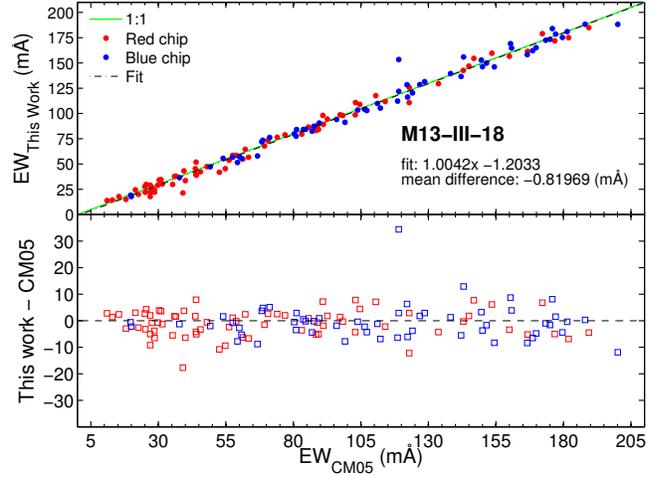}
   \caption{Measured equivalent widths with DAOSPEC vs. CM05 for M13.} 
   \label{fig:EWCohenm13}
\end{figure}

\begin{figure}
   \centering
   \includegraphics[clip=true,trim = 5 0 0 0,width=0.48\textwidth]{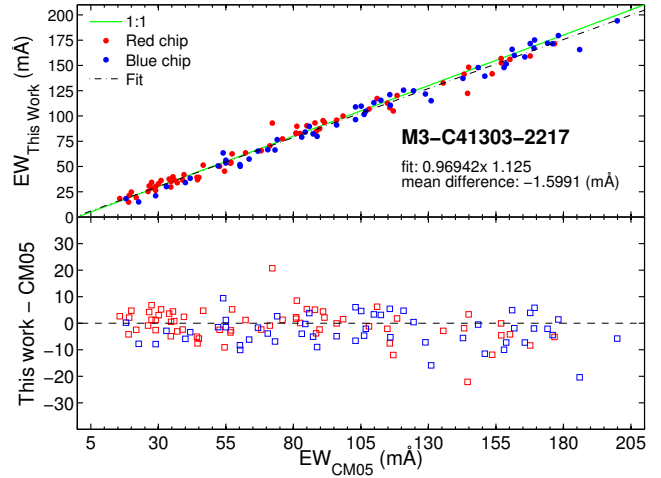}
   \caption{Measured equivalent widths with DAOSPEC vs. CM05 for M3.} 
   \label{fig:EWCohenm3}
\end{figure}

To test the validity of the analysis procedure, two standard stars are analysed using the same method as the target stars: M3-C41303-2217 and M13-III-18. Fig.s \ref{fig:EWCohenm13} and \ref{fig:EWCohenm3} show a comparison of EW measurements (this work) compared with those from \citet[hereafter referred to as CM05]{b2}. A mean difference (This Study - Literature) of $-0.819$ m\AA \hspace{0.20 mm} ($\sigma = 5.836$ m\AA) is found in M13-III-18 (from 127 lines) and a difference of $-1.599$ m\AA\hspace{0.20 mm} ($\sigma = 5.967$ m\AA) is found for M3-C41303-2217 (from 115 lines). This agreement is reasonable and these measurement methods are applied to the new targets in NGC 5024 and NGC 5466.

\section{Model Atmosphere and Abundance Analysis of Optical Data}
\label{modelatmospheres}
The optical abundances are computed by using the radiative transfer code MOOG (\citealt{b8}), which uses stellar model atmospheres and atomic data to synthesize spectra. The new MARCS spherical models have been adapted (\citealt{b3,b79}, also see \citealt{b80}) assuming [$\alpha$/Fe] enhancement since the science targets and standard stars are relatively metal-poor ([Fe/H]~$<$~-1 dex). This combination of spherical model atmospheres and MOOG's plane parallel radiative transfer is appropriate for analyzing red giants (e.g. see \citealt{b76} and additional tests by \citealt{b77}).

\subsection{Photometric Stellar parameters}
\label{sec:stellarparams}
The stellar parameter $\mathrm{T}_{\mathrm{eff}}$ is computed using the infrared flux method (IRFM, described by \citealt{b6}), which requires precision photometry, reddening, distance modulus, metallicity, and stellar mass. Using this $\mathrm{T}_{\mathrm{eff}}$ and an estimate of the stellar mass, the surface gravity can be determined.

Optical photometry comes from the Stetson database\footnote{http://www4.cadc-ccda.hia-iha.nrc-cnrc.gc.ca/community/STETSON/standards/} and the infrared photometry comes from 2MASS \citep[see Table \ref{table:phot}]{b22}. Typical errors for both photometric sources are $\sim0.02$ mag. Reddening values are from the Harris catalogue; typical reddening errors are 10\%. Distance moduli are from the Harris catalogue as determined from the horizontal branch magnitude (taken from \citealt{b23} for NGC 5024 and from \citealt{b24} for NGC 5466). The uncertainty in the distance moduli are 0.1 mag. Input metallicities are taken from the literature: either from direct Fe measurements or from the average cluster metallicity (see Table \ref{table:phot}). Assuming the RGB mass is roughly equivalent to the main sequence turnoff (MSTO) mass, an input stellar mass of $0.8M_{\odot}$ is taken from the MSTO of the isochrone fits of NGC 5024/5466 \citep{b25}\footnote{the actual isochrones corresponding to the GC parameters in \cite{b25} can be downloaded from The Dartmouth Stellar Evolution Database (http://stellar.dartmouth.edu/models/) and manually searched for a turn-off mass.}. The isochrone with the larger age uncertainty is NGC 5466 ($\sigma = 0.75$ Gyr, reflecting a larger intrinsic scatter in its CMD), resulting in a maximum turnoff mass uncertainty of $\Delta M_\mathrm{{turnoff}}~=~3\%$. 

\begin{table*}
 \centering
\hskip-1.5cm\begin{minipage}{140mm}
\caption{Photometric magnitudes and cluster properties}
\begin{tabular}{@{}lccccccccr@{}}
  \hline
& \multicolumn{6}{c}{Photometry\footnote{All B, V, and I values taken from the Stetson database \citep{b9}. All J, H, K values taken from the 2MASS survey.}} & \multicolumn{3}{c}{Cluster parameters\footnote{Distance modulus and reddening taken from \citet{b7}}$^,$\footnote{NGC 5024/5466 metallicity taken from the cluster metallicity itself \citep{b7}, M3 and M13 metallicities taken from CM05. These metallicities are used as input to determine initial stellar parameters.}} \\ \cmidrule(r){2-7} \cmidrule(r){8-10}
 Star  & B & V & I & J & H & K & $(m-M)_V$	& $E(B-V)$ 	& Initial [Fe/H] (dex)  \\
    \hline
NGC 5024-22254 	&15.92	&14.88	&13.74	&12.932	&12.367	&12.230	&16.32		&0.02		&-2.10\\
NGC 5024-50371 	&15.64	&14.56	&13.35	&12.574	&11.973	&11.878	&16.32		&0.02		&-2.10\\
NGC 5466-9951 	&15.90	&14.97	&13.92	&13.092	&12.551	&12.472	&16.02		&0.00		&-1.98\\
NGC 5466-1344 	&15.67	&14.67	&13.58	&12.725	&12.153	&12.056	&16.02		&0.00		&-1.98\\
NGC 5466-10186  	&15.56	&14.62	&13.55	&12.712	&12.137	&12.107	&16.02		&0.00		&-1.98\\
M3-C41303-2217  	&14.86	&13.75	&12.58	&11.698	&11.118	&10.969	&15.07		&0.01		&-1.37$\pm$0.05\\
M13-III-18  		&13.94	&12.74	&11.51	&10.601	&9.928	&9.830	&14.33		&0.02		&-1.43$\pm$0.05\\
\hline
 \label{table:phot}
\end{tabular}
\end{minipage}
\end{table*}

\subsection{Spectroscopic Stellar Parameters}
\label{sec:aspcapref}

\textcolor{black}{Stellar model atmospheres were interpolated from the grid of MARCS models using the initial parameters discussed in Section \ref{sec:stellarparams}}. These models were run with MOOG to compute initial line abundances from EW measurements. Revised values of $\mathrm{T}_{\mathrm{eff}}$ and \textcolor{black}{$\log\epsilon$(\ion{Fe}{1})}, and new microturbulence values were found. $\mathrm{T}_{\mathrm{eff}}$ is adjusted in small increments to find the best fit to the \ion{Fe}{1} line abundances when plotted against the excitation potential ($\chi$). Simultaneously the microturbulence is adjusted such that there is no dependence between $\log\epsilon$(\ion{Fe}{1}) and the reduced EWs. This process is repeated until the final slopes in \ion{Fe}{1} vs $\chi$ and reduced EW were $<$ 0.004. These spectroscopic parameters are listed in Table \ref{table:inputParam}\textcolor{black}{; in the case of $\mathrm{T}_{\mathrm{eff}}$ the spectroscopic and photometric values agree within the errors. We refer the reader to Section \ref{sec:absens} to demonstrate how these $\mathrm{T}_{\mathrm{eff}}$ differences affect the derived stellar abundances (i.e. see Table \ref{table:absens}).} Also included in the table are the derived stellar parameters from the APOGEE Stellar Parameters and Chemical Abundance Pipeline (ASPCAP\footnote{ASPCAP parameters were retrieved online at http://data.sdss3.org/sas/dr10/apogee/spectro/redux/r3/fields.html}, Garc\'ia P\'erez et al. 2014, in preparation) for Fe, C and N.

\begin{table*}
\centering
\hskip-3.5cm\begin{minipage}{140mm}
\caption{Derived Temperatures and Gravity}
\begin{tabular}{@{}lcccccccccc@{}}
\hline
& \multicolumn{4}{c}{This Study} & \multicolumn{3}{c}{CM05} & \multicolumn{3}{c}{ASPCAP\footnote{The uncertainties in these stellar parameters are very small and are therefore not included here.} (APOGEE)} \\ \cmidrule(r){2-5} \cmidrule(r){6-8}  \cmidrule(r){9-11}
Star           			&  \multicolumn{2}{c}{$\mathrm{T}_{\mathrm{eff}}$ ($\pm$100 K)}\footnote{\textcolor{black}{In this work we adopt spectroscopic $\mathrm{T}_{\mathrm{eff}}$ as there are typically $>$ 100 \ion{Fe}{1} lines available to constrain this parameter.}} &  $\log(g)$	& $v_t$		&$\mathrm{T}_{\mathrm{eff}}$&  $\log(g)$	& $v_t$	&$\mathrm{T}_{\mathrm{eff}}$&  $\log(g)$	& $v_t$\\
				&\textcolor{black}{Phot.} &Spect.& ($\pm$0.2 dex)& ($\pm$0.2 km s$^{-1}$)& ($\pm$75 K)& ($\pm$0.2 dex)& ($^{+0.4}_{-0.2}$km s$^{-1}$)& (K)	&  dex)		& (km s$^{-1}$)\\
\hline
NGC 5024-22254 	&4511&4410		&1.21		&1.60		&-			&-			&-		&4409		&1.37		&1.83		\\
NGC 5024-50371 	&4444&4425		&1.06		&1.80		&-			&-			&-		&4391		&1.26		&1.86		\\
NGC 5466-9951 	&4595&4600		&1.44		&1.55		&-			&-			&-		&4512		&1.75		&1.71		\\
NGC 5466-1344 	&4512&4499		&1.27		&1.40		&-			&-			&-		&4469		&1.62		&1.75		\\
NGC 5466-10186  	&4571&4585		&1.29		&1.75		&-			&-			&-		&-			&-			&-			\\
M3-C41303-2217  	&4379&4538		&1.19		&1.55		&4436 		&1.20		&1.60	&4319		&1.69		&1.73		\\
M13-III-18  		&4298&4397		&1.02		&1.75		&4350		&1.00		&1.65	&4498		&2.25		&1.56		\\
\hline \\[0.10cm]
\label{table:inputParam}
\end{tabular}
\end{minipage}
\end{table*}

\subsection{Stellar Parameter Uncertainties}

\label{sec:teff}
The uncertainties in $\mathrm{T}_{\mathrm{eff}}$ were found by assuming the derived individual \ion{Fe}{1} line abundances are distributed randomly about the mean; thus the standard deviation ($\sigma_{\mathrm{Fe}}$) in the mean abundance measures the scatter in the data. If $\mathrm{T}_{\mathrm{eff}}$ is adjusted such that the slope of the log(Fe) vs $\chi$ spans $\pm 1 \sigma$ over the range in excitation potentials ($\chi$), then $\Delta\mathrm{T}_{\mathrm{eff}}$ is found to be $\sim$100 K for both standard stars, and this is adopted for the five science targets given the similarity in S/N and $\mathrm{T}_{\mathrm{eff}}$ in our sample. 

The $\Delta$$v_t$ is found by allowing the slope log$\epsilon$(Fe I) vs. reduced EW to vary such that the range of log$\epsilon$(\ion{Fe}{1}) values span 1 $\sigma$ over the range of reduced EWs, similar to the method used to find $\Delta\mathrm{T}_{\mathrm{eff}}$. The $\Delta$$v_t = 0.2$ $\mathrm{kms^{-1}}$ for both standard stars and is adopted for the five science targets.

Physical gravities are adopted from the IRFM, thus log(g) depends on the distance to the star (or cluster), reddening, $\mathrm{T}_{\mathrm{eff}}$ value, turn-off mass, and photometry; an estimate in the uncertainty of log(g) is calculated by quantifying and propagating the errors in these parameters. After calculating log(g) and varying these parameters by the errors described in this Section, $\Delta$log(g) = 0.2 dex (Assuming no covariance).

\subsection{Comparison of Stellar Parameters and Iron with the Standard Stars}
\label{sec:Fecomp}
To test the validity of this abundance analysis technique the procedure is applied to the standard stars in M3 and M13 using the model atmospheres and EWs of this work and comparing the \ion{Fe}{1} abundance with CM05. The model atmosphere parameters of CM05 agree within the uncertainties of this work (see Table \ref{table:inputParam}) and so the Fe comparison should yield consistent results. The derived M13 standard star [Fe/H] abundance from this work is $-1.57 \pm 0.05$ whereas that of CM05 is $-1.48 \pm 0.05$ and that of \citet{b12} is $-1.50 \pm 0.05$ (after correcting both for different solar Fe abundances). The derived M3 standard star [Fe/H] abundance is $-1.33 \pm 0.05$ whereas that of CM05 is $-1.42 \pm 0.05$ (again correcting for different solar Fe abundances). We consider these results in good agreement (see Section \ref{sec:CMcomp} for an abundance comparison of all determined elements).
 
\section{Abundance Analysis of Infrared Data}

\label{sec:iranal}

Chemical abundances in the IR are determined via spectrum synthesis, and not EWs. EWs are not practical in the IR due to the presence of molecular lines and blends across the entire spectrum. Synthetic spectra are calculated with MOOG using the model atmospheres described in Section \ref{modelatmospheres} and the DR10 line list from Matthew Shetrone (Shetrone et al. in prep). \textcolor{black}{A $\mathrm{^{12}}$C/$\mathrm{^{13}}$C ratio of 6 is adopted following \cite{b21}. However, at low metallicities the spectral features sensitive to this ratio (i.e. $^{13}$C$^{16}$O) show no observable change when altering this ratio from 6 to 50. There is also no observable change in overall carbon abundance after altering the ratio by this amount. Therefore the choice of this value over this range is not critical.}

To find the iron abundance a preliminary selection of the most prominent iron lines ($\sim$20) is chosen over the spectral range 15100-16900 \AA\hspace{0.20 mm}. The spectral region around each of these lines is synthesized with an iron abundance of zero to ensure the feature of interest vanishes and that no other elements or molecules contribute to the spectral line. The resulting sample of iron lines is typically reduced to $\sim$12. The uncertainty for each measured iron line is determined by examining the residual between the synthetic and actual spectrum and finding the abundance which causes the residual to differ visually from the noise. The weighted average is computed for all of the individually examined iron lines and then set as the iron abundance for the synthesis of the rest of the elements. The error of the weighted mean is taken to be the final abundance uncertainty; this method is used to compute the uncertainties in all infrared abundance calculations except in the case of oxygen.

\begin{figure}
   \centering
    \includegraphics[clip=true,trim = 5 0 0 0,width=0.48\textwidth]{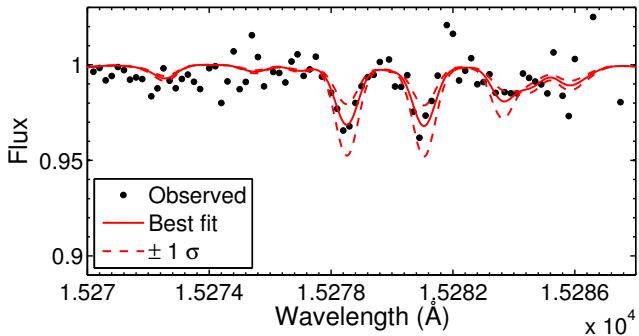}
   \caption{Example of a typical OH line measured in the infrared. This particular \textcolor{black}{case} is taken from NGC 5466 1344. The two dotted lines represent 1 $\sigma$ abundance errors.}
   \label{fig:OHline}
\end{figure}

Oxygen has a slight dependance on the carbon abundance through CO formation\textcolor{black}{. Therefore} an initial carbon abundance is determined from the analysis of a single atomic \ion{C}{1} line and several CO features, where carbon is much more sensitive than oxygen. With carbon fixed, the oxygen is then determined by synthesizing OH lines (see Fig. \ref{fig:OHline}).ÊThe new oxygen abundance is fixed, and the carbon redetermined. This iteration is repeated until the carbon and oxygen abundances converge (typically two iterations). The oxygen uncertainty is first determined exactly as described above with iron, then it is added in quadrature to the value found from re-deriving the oxygen abundance after fluctuating carbon by its maximum error. Finally, nitrogen is found from CN regions using the final carbon abundance (see section \ref{sec:carbon} for the regions). The rest of the elements are then synthesized and their abundances determined. The final abundances are shown in Tables \ref{table:Ab1} and \ref{table:Ab2}.

\section{Abundance Results}

\label{sec:hfs}

In this section, the abundance errors and standard star comparison are first addressed before comparing our results with the Galactic sample. The abundances with respect to \ion{Fe}{1} from both wavelength regions are then compared with the literature in Section \ref{sec:galcomp}. The reported abundances are with respect to solar, (\cite{b20}). Galactic comparison stars are taken from \cite{b27}, \cite{b28} (metal-poor stars), and \cite{b44} (thick disk stars) and comparison MW clusters are taken from \cite{b45}. Additional comparison stars for Cu and Zn are included from \cite{b29}. These comparison stars were chosen because their abundances were determined from high-resolution, high S/N spectra. For each element the specific spectral lines used to compute the abundances are discussed as well as any additional effects or corrections (i.e. hyperfine structure, NLTE, 3D effects, etc.). Hyperfine structure corrections are applied to the elements Sc, V, Mn, Co, Cu, Ba, and La; the isotopic data used to compute the corrections were collected from a variety of sources (Sc, V, Mn, and Co from \citealt{b63}, additional Mn from \citealt{b64}, Cu from \citealt{Biehl}, Ba from \citealt{b66}, Eu and La from \citealt{b67}), and the corrections are only found to be significant ($>$0.05 dex) for V and Mn. The abundance errors and standard star comparisons are also discussed in this section.

\subsection{Abundance Errors}
\label{sec:absens}
The abundance errors are determined by combining intrinsic random errors in the sample with the errors associated with the stellar parameter uncertainties. The intrinsic errors in the sample are determined by computing the error in the mean (i.e. $\sigma_X$~/~$\sqrt{N_\mathrm{X}}$) except in the case where there were fewer than 5 lines of element X, in which case $\sigma_{\mathrm{Fe }}$~/~$\sqrt{N_\mathrm{X}}$ is used. The abundance sensitivities due to stellar parameter uncertainties are calculated and summarized for a sample star (NGC 5024-22254) in Table \ref{table:absens}. The final errors used with the Galactic comparison are determined by adding the random and stellar parameter uncertainties in quadrature (ignoring the covariant terms between temperature and microturbulence).

\begin{table}
	\centering
	\caption{Abundance Sensitivities for NGC 5024-22254}
	\setlength{\tabcolsep}{6.5pt}
	\def\arraystretch{0.90}

		\begin{minipage}{85mm}
			\label{table:absens}
			\begin{tabular}{@{}lcccr@{}}
			\hline
			\hline
				Species	& $\Delta T_{\mathrm{eff}}$ &$\Delta$ log g	& $\Delta v_t$		&Total\footnote{All errors added in quadrature.}   \\
						&(+ 100K)				   &(+ 0.2 dex)	&(+ 0.2 km/s)	&						\\
				\hline
				\multicolumn{5}{c}{Optical Data}\\
				\hline
				
				\ion{Fe}{1}		&-0.15				&0.00		& 0.10		&0.18					\\
				\ion{Fe}{2}		&0.06				&-0.07		& 0.05		&0.10					\\
				\ion{O}{1}			&-0.01				&-0.04		& 0.01		& 0.04					\\
				\ion{Na}{1}		&-0.15				&0.04		& 0.06		&0.17					\\
				\ion{Mg}{1}		&-0.09				&0.03		& 0.05		&0.11					\\
				\ion{Al}{1}			&...					&...			&...			&...						\\
				\ion{Si}{1}			&-0.01				&-0.01		& 0.01		&0.02					\\
				\ion{Ca}{1}		&-0.12				&0.01		& 0.05		&0.13					\\
				\ion{Sc}{2}		&0.01				&-0.07		& 0.04		&0.08					\\
				\ion{Ti}{1}			&-0.21				&-0.01		& 0.05		&0.22					\\
				\ion{Ti}{2}			&0.01				&-0.06		& 0.08		&0.10					\\
				\ion{V}{1}			&-0.21				&-0.01		& 0.00		&0.21					\\
				\ion{Cr}{1}			&-0.21				&-0.01		& 0.08		&0.22					\\
				\ion{Cr}{2}			&0.06				&-0.06		& 0.03		&0.09					\\
				\ion{Mn}{1}		&-0.15				&0.02		& 0.08		&0.17					\\
				\ion{Co}{1}		&-0.15				&-0.01		& 0.01		&0.15					\\
				\ion{Ni}{1}			&-0.13				&-0.01		& 0.03		&0.13					\\
				\ion{Cu}{1}		&-0.15				&-0.01		& 0.01		&0.01					\\
				\ion{Zn}{1}		&0.04				&-0.03		& 0.04		&0.06					\\
				\ion{Y}{2}			&-0.01				&-0.07		& 0.10		&0.10					\\
				\ion{Ba}{2}		&-0.05				&-0.07		& 0.12		&0.15					\\
				\ion{La}{2}		&-0.03				&-0.08		& 0.01		&0.09					\\
				\ion{Nd}{2}		&-0.03				&-0.08		& 0.01		&0.09					\\
				\ion{Eu}{2}		&...				&...			& ...			&						\\
				\hline
				\multicolumn{5}{c}{Infrared Data}\\
				\hline
				\ion{Fe}{1}		&-0.07				& 0.01		& 0.00 		& 0.07					\\
				\ion{C}{1}			&-0.01				&-0.08		&-0.05		& 0.09					\\
				\ion{N}{1}			&-0.15				& 0.09		& 0.06		& 0.18					\\
				\ion{O}{1}			&-0.18				& 0.00		&-0.01		& 0.18					\\
				\ion{Mg}{1}		&-0.13				&-0.01		&-0.02		& 0.14					\\
				\ion{Al}{1}			&-0.11				&-0.01		&-0.04		& 0.12					\\
				\ion{Si}{1}			&-0.12				&-0.03		&-0.03		& 0.12					\\
				\ion{Ca}{1}		&-0.19				&-0.10		&-0.14		& 0.26					\\
				\ion{Ti}{1}			&-0.20				&-0.05		&-0.05		& 0.21					\\
				\hline
			\end{tabular}
			\end{minipage}
\end{table}

\subsection{Standard Star Comparison}
\label{sec:CMcomp}

The standard star abundances are compared to CM05 in detail here. The EWs and stellar parameters between this work and CM05 are in good agreement, as summarized in Sections \ref{sec:EWcomp} and \ref{sec:Fecomp}; thus the remaining elemental abundances are now compared. The standard star abundances are shown in Table \ref{table:CohenComp} \textcolor{black}{along with their random errors (the systematic errors are not included in the table to keep consistent with the similar table found in CM05)}; the average abundance difference ($\pm \sigma$) between this work and CM05 is 0.08 $\pm$ 0.13 and 0.14 $\pm$ 0.11 for M3-C41303-2217 and M13-III-18 respectively which we consider good agreement. In the case where there are derived abundances for both the neutral and ionized species (except for the case of Fe), the weighted mean and error between the two is reported and compared with CM05. As opposed to the other elemental abundances, oxygen was determined with respect to \ion{Fe}{2} in order to keep consistent with CM05. 

For the M3 star most of the elements agree within the errors with the exception of Sc, V, Ni, and La (which lie within 2 $\sigma$ of each other), of Ca (3 $\sigma$), and of Zn (4 $\sigma$). \cite{b12} find Sc to be -0.11 dex lower than CM05 and we find also find Sc to be lower (by -0.16 dex). The V abundance in this work has only 3 overlapping lines with CM05, all with different log(gf) values ($\sim$0.05 dex difference). The Ni abundances in this work are derived with log(gf) values that are quite different than CM05, with all 8 of the overlapping lines showing differences up to 0.24 dex. La is derived from 2 lines in this work with similar EWs and log(gf) values to CM05, and from one additional La line; after taking into account the differences in adopted solar values between this work and CM05 the La abundances agree within the errors. The M3 Ca abundance derived by \cite{b12} differs from CM05 by -0.13 dex, and the difference between this work and CM05 is also found to be lower (by -0.17 dex). There is only one common Zn line between this work and CM05; this line differs in both EW (by 17$\%$) and in log(gf) value (by 0.05 dex). \textcolor{black}{The abundances of all elements mentioned here agree with CM05 when taking into account the systematic uncertainties, except in the case of Zn.}

The M13 star also shows most elements agree within the errors with the exception of Na, and V (which lie within 2 $\sigma$), of Ti, Ni, and Y (3 $\sigma$), and of Al, and Ca (4 $\sigma$). The Na abundances are computed from log(gf) values that differ with CM05 by up to 0.05 dex and with excitation potentials that differ by up to 0.10 ev. V, Ni, Ca, and La derived abundances share the same discrepancies discussed above with the M3 standard star. The Ti abundance is calculated from an average between TiI and TiII and in this work there are many more TiII lines than with CM05. The NLTE effects on TiI are well known and result in a lower TiI abundance compared to TiII (see Section \ref{sec:tidiff}), thus an average computed with more TiI lines than TiII (such as CM05) will inevitably ensure an overall lower Ti abundance as is shown for both standard stars. The Y abundance is derived from 4 lines, 3 of which are in common with CM05; the line that is not in common with CM05 yields a higher abundance by $\sim$ 0.2 dex. The Al abundance is derived from entirely different lines than CM05. For La there are only two out of four lines that overlap with CM05 and there are $\sim 15 \%$ discrepancies in EW for the common lines; furthermore if [La/\ion{Fe}{1}] is calculated using the solar values and \ion{Fe}{1} parameters of CM05 then the La abundances agree. \textcolor{black}{The comparisons of all elements mentioned here agree with CM05 when taking into account the systematic uncertainties, except in the case of Al as there is no reported systematic for this element in CM05.}

Overall, the standard star comparison is within good agreement and we consider the methods used both here and with the target stars sound. The target star abundances are discussed in detail in the following section.

\begin{table}
	\centering
	\caption{Standard star abundance comparison}
	\setlength{\tabcolsep}{1pt}
	\def\arraystretch{0.90}

		\begin{minipage}{85mm}
			\label{table:CohenComp}
			\begin{tabular}{@{}lccccccr@{}}
			\multicolumn{8}{c}{M3-C41303-2217 comparison with CM05}\\
			\hline
			\hline
				Species\footnote{A line weighted average is calculated for species with more than one ionization state to stay consistent with CM05. The [Ab/Fe] is calculated with respect to \ion{Fe}{1} except for with O which is calculated with \ion{Fe}{2} to match CM05.}			& Abundance	&$\sigma$	& N		& Abundance									&$\sigma$		& N 		& $\Delta$\footnote{$\Delta$ = (This Work - CM05)}	   \\
				\hline
								 \multicolumn{3}{r}{This work}  & \multicolumn{3}{r}{CM05}		&					 \\
				\hline
				
				{[}\ion{Fe}{1}/H]		&-1.33 		&0.05		& 103	&-1.37			&0.05			&126	&0.04		\\
				{[}\ion{Fe}{2}/H]		&-1.33 		&0.05		& 14		&-1.34			&0.05			&14		&0.01		\\
				{[}O/Fe]			& 0.49		&0.10		& 2		& 0.33			&0.05			&2		&0.10		\\
				{[}Na/Fe]			&-0.27		&0.05		& 2		&-0.39			&0.09			&3		&0.12		\\
				{[}Mg/Fe]			&0.33		&0.13		& 2		&0.40			&0.11			&3		&-0.07		\\
				{[}Al/Fe]			&0.32		&0.20		& 2		&...				&...				&...		&...			\\
				{[}Si/Fe]			&0.16		&0.08		& 2		&0.20			&0.05			&9		&-0.04		\\
				{[}Ca/Fe]			&0.25		&0.05		& 16		&0.08			&0.05			&15		&0.17		\\
				{[}Sc/Fe]			&-0.01		&0.05		& 3		&0.15			&0.05			&7		&-0.16		\\
				{[}Ti/Fe]			&0.17		&0.05		& 29		&0.10			&0.05			&25		&0.07		\\
				{[}V/Fe]			&0.05		&0.10		& 12		&-0.18			&0.06			&8		&0.23		\\
				{[}Cr/Fe]			&0.08		&0.05		& 7		&0.01			&0.06			&9		&0.07		\\
				{[}Mn/Fe]			&-0.25		&0.13		& 6		&-0.27			&0.14			&4		&0.02		\\
				{[}Co/Fe]			&0.10		&0.30		& 2		&-0.11			&0.03			&3		&0.21		\\
				{[}Ni/Fe]			&0.03		&0.08		& 16		&-0.16			&0.05			&15		&0.19		\\
				{[}Cu/Fe]			&-0.44		&0.09		& 3		&-0.56			&0.21			&2		&0.12		\\
				{[}Zn/Fe]			&0.24		&0.08		& 2		&-0.10			&0.05			&2		&0.34		\\
				{[}Y/Fe]			&-0.21		&0.08		& 4		&-0.15			&0.10			&4		&-0.06		\\
				{[}Ba/Fe]			&0.12		&0.08		& 4		&0.16			&0.05			&3		&-0.04		\\
				{[}La/Fe]			&0.23		&0.05		& 3		&-0.02			&0.15			&2		&0.25		\\
				{[}Nd/Fe]			&0.25		&0.07		& 4		&0.28			&0.06			&5		&-0.03		\\
				{[}Eu/Fe]			&0.65		&0.10		& 1		&0.51			&0.10			&1		&0.14		\\
				\hline
								&			&			&		\multicolumn{3}{c}{mean $\pm$  st. dev.} 	&	$\overline{\Delta} =$& $0.08\pm0.13$ \\
				\hline
				\hline
				\multicolumn{8}{c}{M13-III-18 comparison with CM05}\\

				\hline

				{[}\ion{Fe}{1}/H]			&-1.57 		&0.05		& 118	&-1.43			&0.05			&123	&-0.14		\\
				{[}\ion{Fe}{2}/H]			&-1.34 		&0.05		& 23		&-1.46			&0.05			&13		&0.12		\\
				{[}O/Fe]			& ...			&...			& ...		& 0.35			&0.10			&1		&...			\\
				{[}Na/Fe]			&0.57		&0.08		& 4		&0.36			&0.06			&4		&0.21		\\
				{[}Mg/Fe]			&0.39		&0.13		& 3		&0.29			&0.14			&3		&0.10		\\
				{[}Al/Fe]			&1.07		&0.06		& 2		&0.74			&0.10			&1		&0.33		\\
				{[}Si/Fe]			&0.41		&0.05		& 6		&0.36			&0.05			&13		&0.05		\\
				{[}Ca/Fe]			&0.32		&0.05		& 15		&0.10			&0.05			&13		&0.22		\\
				{[}Sc/Fe]			&0.31		&0.20		& 5		&0.21			&0.05			&7		&0.10		\\
				{[}Ti/Fe]			&0.27		&0.05		& 46		&0.09			&0.05			&22		&0.18		\\
				{[}V/Fe]			&0.04		&0.07		& 11		&-0.11			&0.05			&8		&0.15		\\
				{[}Cr/Fe]			&0.11		&0.07		& 8		&-0.01			&0.05			&7		&0.12		\\
				{[}Mn/Fe]			&-0.25		&0.09		& 7		&-0.25			&0.08			&4		&0.00		\\
				{[}Co/Fe]			&0.16		&0.27		& 2		&-0.02			&0.06			&4		&0.18		\\
				{[}Ni/Fe]			&0.08		&0.05		& 11		&-0.09			&0.05			&18		&0.17		\\
				{[}Cu/Fe]			&-0.45		&0.08		& 3		&-0.61			&0.13			&2		&0.16		\\
				{[}Zn/Fe]			&0.09		&0.10		& 1		&-0.05			&0.05			&2		&0.14		\\
				{[}Y/Fe]			&0.01		&0.09		& 5		&-0.26			&0.06			&5		&0.27		\\
				{[}Ba/Fe]			&0.39		&0.05		& 4		&0.36			&0.05			&3		&0.03		\\
				{[}La/Fe]			&0.41		&0.06		& 4		&0.09			&0.09			&3		&0.32		\\
				{[}Nd/Fe]			&0.40		&0.05		& 3		&0.31			&0.06			&7		&0.09		\\
				{[}Eu/Fe]			&0.70		&0.10		& 1		&0.58			&0.10			&1		&0.12		\\
				\hline
								&			&			&		\multicolumn{3}{c}{mean $\pm$  st. dev.} 	&	$\overline{\Delta} =$& $0.14\pm0.11$ \\
				\hline

			\end{tabular}
			\end{minipage}
			
\end{table}

\subsection{NGC 5024/5466 Stars}
\label{sec:galcomp}

\textcolor{black}{The final optical and infrared stellar abundances of the target stars are reported in Tables \ref{table:Ab1}, \ref{table:Ab2}, and \ref{table:Ab3}, along with their random errors. For all of the comparisons with the Galactic sample (i.e. Figures 6-14), the abundance uncertainties reflect both the errors from the stellar parameters and random errors reported in the aforementioned Tables; the two errors are added in quadrature.}

\begin{table*}
\centering
\begin{minipage}{230mm}
\caption{Derived abundances for NGC5466 and NGC5024: \ion{Fe}{1}, C, N, O}
\begin{tabular}{@{}lcccccc@{}}
\hline
		&Data			&\multicolumn{5}{c}{Abundance ([X/Fe])\footnote{[X/H] is given for \ion{Fe}{1} instead.} $\pm\sigma /  \sqrt{N}$ (\#)\footnote{For elements where N $<$ 5 and $\sigma$ is less than $\sigma_{\mathrm{Fe I}}$, the error is reported as $\pm\sigma_{\mathrm{Fe I}}/\sqrt{N}$; this is also applied to Tables \ref{table:Ab2} and \ref{table:Ab3}. }$^,$\footnote{A * indicates the case where there was a synthetic fit \textcolor{black}{to molecular bands of $^{12}$C$^{16}$O and $^{12}$C$^{14}$N.}}	 }\\
Species\footnote{All abundances ratios computed with respect to NLTE corrected \ion{Fe}{1}.}	&source		&NGC5024-22254		&NGC5024-50371		&NGC5466-9951		&NGC5466-1344		&NGC5466-10186		\\

\hline
\ion{Fe}{1}&Optical		&$-2.17\pm0.05$ (105)	&$-2.15\pm0.05$ (114)	&$-1.98\pm0.05$ (114)	&$-2.02\pm0.05$ (114)	&$-2.19\pm0.05$ (110)	\\
 		&Infrared		&$-2.19\pm0.07$ (12)	&$-2.12\pm0.07$ (12)	&$-2.00\pm0.07$ (16)	&$-2.06\pm0.06$ (11)	&...					\\
		&Ave. NLTE\footnote{Average [\ion{Fe}{1}/H] between optical and infrared (where applicable), corrected for NLTE.}	&$-2.08\pm0.09$	&$-2.04\pm0.09$ 	&$-1.89\pm0.09$ 	&$-1.94\pm0.08$ 	&$-2.09\pm0.05$ 				\\
 		&ASPCAP	&$-2.07\pm0.01$ 		&$-1.99\pm0.01$		&$-1.89\pm0.05$ 		&$-1.94\pm0.01$ 		&...					\\
		
\ion{C}{1}	&Optical		&...					&...					&...					&...					&...					\\
 		&Infrared		&$-0.12\pm0.37$ (*)		&$-0.28\pm0.33$ (*)		&$-0.33\pm0.35$ (*)		&$-0.20\pm0.20$ (*)		&...					\\
 		&ASPCAP	&$-0.02\pm0.03$ 		&$-0.15\pm0.05$		&$-0.35\pm0.17$ 		&$-0.17\pm0.08$ 		&...					\\

\ion{N}{1}	&Optical		&...					&...					&...					&...					&...					\\
 		&Infrared		&$0.21\pm0.32$ (*)		&$0.46\pm0.38$ (*)		&$0.70\pm0.30$ (*)		&$0.58\pm0.21$ (*)		&...					\\
 		&ASPCAP	&$0.49\pm0.04$ 		&$0.57\pm0.05$		&$0.87\pm0.17$ 		&$0.76\pm0.10$ 		&...					\\

\ion{O}{1}	&Optical		&$0.74\pm0.18$ (1)		&$0.89\pm0.17$ (1)		&$0.38\pm0.19$ (1)		&$0.59\pm0.16$ (1)		&$<0.40$				\\
 		&Infrared		&$0.16\pm0.12$ (12)	&$0.29\pm0.11$ (12)	&$0.32\pm0.12$ (12)	&$0.27\pm0.11$ (12)	&...					\\
 		&ASPCAP	&...					&...					&...			 		&...					&...					\\

\hline
\label{table:Ab1}
\end{tabular}

\end{minipage}

\end{table*}

\begin{table*}
\centering
\begin{minipage}{230mm}
\caption{Derived abundances for NGC5466 and NGC5024: Elements in common between Optical and IR}
\begin{tabular}{@{}lcccccc@{}}
\hline
		&Data			&\multicolumn{5}{c}{Abundance ([X/Fe])$\pm\sigma / \sqrt{N} $ (\#)} 	\\
Species\footnote{All abundances ratios computed with respect to NLTE corrected \ion{Fe}{1}.}			&source		&NGC5024-22254		&NGC5024-50371		&NGC5466-9951		&NGC5466-1344		&NGC5466-10186	\\

\hline

\ion{Mg}{1}		&Optical		&$0.16\pm0.27$ (3)		&$0.50\pm0.14$ (3)		&$0.24\pm0.14$ (3)		&$0.27\pm0.15$ (3)		&$0.32\pm0.13$ (3)	\\
 				&Infrared		&$0.23\pm0.12$ (3)		&$0.25\pm0.12$ (3)		&$-0.09\pm0.16$ (3)		&$0.09\pm0.13$ (3)	&-					\\
		
\ion{Al}{1}\footnote{\textcolor{black}{The optical \ion{Al}{1} placeholder is included here to demonstrate this element \textit{can} be derived from both the optical and the infrared.}}		&Optical		&...					&...					&...					&...					&...				\\
 				&Infrared		&$-0.03\pm0.14$ (2)		&$-0.18\pm0.21$ (2)		&$0.25\pm0.16$ (3)		&$-0.36\pm0.21$ (2)		&...				\\

\ion{Si}{1}			&Optical		&$0.68\pm0.18$ (1)		&$0.34\pm0.17$ (1)		&$0.34\pm0.14$ (1)		&$0.36\pm0.16$ (1)		&$0.41\pm0.17$ (1)	\\
 				&Infrared		&$0.23\pm0.06$ (11)	&$0.34\pm0.07$ (11)	&$0.12\pm0.08$ (11)	&$0.12\pm0.08$ (11)	&...				\\

\ion{Ca}{1}		&Optical		&$0.31\pm0.11$ (14)	&$0.26\pm0.07$ (15)	&$0.17\pm0.08$ (17)	&$0.18\pm0.07$ (15)	&$0.19\pm0.07$ (14)\\
 				&Infrared		&$0.23\pm0.19$ (2)		&$0.25\pm0.42$ (2)		&$0.12\pm0.31$ (2)		&$0.15\pm0.25$ (1)		&...				\\
		
\ion{Ti}{1}		 	&Optical		&$-0.14\pm0.09$ (17)	&$-0.09\pm0.12$ (17)	&$-0.24\pm0.07$ (15)	&$0.12\pm0.11$ (16)	&$0.13\pm0.11$ (14)\\
 				&Infrared		&$0.15\pm0.20$ (1)		&$<0.30$ 				&$0.15\pm0.28$ (2)		&$<0.40$				&...				\\
\hline
\label{table:Ab2}
\end{tabular}
\end{minipage}
\end{table*}

\begin{table*}
\centering
\begin{minipage}{230mm}
\caption{Derived abundances for NGC5466 and NGC5024: Additional elements from optical data}
\begin{tabular}{@{}lcccccc@{}}
\hline
		& 			&\multicolumn{5}{c}{Abundance ([X/Fe])\footnote{[X/H] is given for \ion{Fe}{2} instead.} $\pm\sigma /  \sqrt{N}$ (\#)} 	\\
Species\footnote{All abundances ratios computed with respect to NLTE corrected \ion{Fe}{1}.}				&Correction		&NGC5024-22254		&NGC5024-50371		&NGC5466-9951		&NGC5466-1344		&NGC5466-10186	\\

\hline
\ion{Fe}{2}				&None		&$-1.88\pm0.13$ (18)	&$-1.90\pm0.09$ (23)	&$-1.65\pm0.05$ (28)	&$-1.84\pm0.10$ (18)	&$-1.89\pm0.08$ (24)\\

\ion{Na}{1}				&None		&$-0.38\pm0.11$ (3)		& $-0.58\pm0.17$ (2)	&$-0.28\pm0.19$ (1)		&$-0.67\pm0.16$ (1)		&$-0.65\pm0.12$ (2)\\
	
\ion{Na}{1}				&NLTE		&$-0.38\pm0.11$ (3)		& $-0.58\pm0.17$ (2)	&$-0.28\pm0.19$ (1)		&$-0.67\pm0.16$ (1)		&$-0.65\pm0.12$ (2)\\
		
\ion{Sc}{2}				&None		&$0.03\pm0.14$ (5)		&$0.33\pm0.14$ (4)		&$-0.01\pm0.08$ (6)		&$0.19\pm0.17$ (5)		&$-0.01\pm0.11$ (3)\\

\ion{Ti}{2}					&None		&$0.52\pm0.09$ (17)	&$0.56\pm0.10$ (20)	&$0.29\pm0.09$ (34)	&$0.37\pm0.09$ (18)	&$0.40\pm0.07$ (23)\\

\ion{V}{1}					&None		&$0.07\pm0.14$ (5)		&$0.27\pm0.17$ (1)		&$-0.10\pm0.19$ (1)		&...					&$0.41\pm0.17$ (1)\\
	
\ion{V}{1}					&HFS		&$0.14\pm0.09$ (5)		&$0.22\pm0.17$ (1)		&$-0.17\pm0.19$ (1)		&...					&$0.34\pm0.17$ (1)\\	
		
\ion{Cr}{1}					&None		&$-0.79\pm0.33$ (5)		&$-0.33\pm0.11$ (3)		&$-0.34\pm0.17$ (5)		&$-0.18\pm0.10$ (3)		&$-0.27\pm0.11$ (3)\\

\ion{Cr}{1}					&NLTE		&$-0.57\pm0.32$ (5)		&$-0.11\pm0.11$ (3)		&$-0.12\pm0.17$ (5)		&$0.04\pm0.10$ (3)		&$-0.05\pm0.11$ (3)\\

\ion{Cr}{2}		 			&None		&$0.50\pm0.25$ (3)		&$0.32\pm0.25$ (3)		&$0.55\pm0.15$ (4)		&$0.36\pm0.18$ (3)		&$0.17\pm0.11$ (4)\\

\ion{Cr}{2}					&NLTE		&$0.55\pm0.25$ (3)		&$0.37\pm0.25$ (3)		&$0.59\pm0.15$ (4)		&$0.40\pm0.18$ (3)		&$0.22\pm0.11$ (4)\\

\ion{Mn}{1}				&None		&$-0.21\pm0.13$ (2)		&$-0.59\pm0.15$ (5)		&$-0.26\pm0.20$ (6)		&$-0.68\pm0.09$ (4)		&$-0.58\pm0.11$ (3)\\

\ion{Mn}{1}			 	&HFS		&$-0.39\pm0.13$ (2)		&$-0.65\pm0.14$ (5)		&$-0.36\pm0.20$ (6)		&$-0.24\pm0.09$ (4)		&$-0.52\pm0.11$ (3)\\

\ion{Co}{1}				&None		&$0.02\pm0.33$ (2)		&...					&$-0.38\pm0.19$ (1)		&$-0.22\pm0.16$ (1)		&...				\\

\ion{Ni}{1}					&None		&$-0.31\pm0.20$ (6)		&$-0.14\pm0.09$ (6)		&$0.00\pm0.21$ (8)		&$-0.18\pm0.09$ (7)		&$-0.05\pm0.13$ (5)\\

\ion{Cu}{1}				&None		&...					&$-0.52\pm0.17$ (1)		&...					&...					&...				\\

\ion{Zn}{1}				&None		&$0.25\pm0.18$ (1)		&$0.30\pm0.17$ (1)		&$-033\pm0.19$ (1)		&$-0.07\pm0.16$ (1)		&$0.01\pm0.17$ (1)\\

\ion{Y}{2}					&None		&$0.00\pm0.21$ (4)		&$-0.11\pm0.13$ (4)		&$-0.43\pm0.12$ (3)		&$-0.48\pm0.10$ (3)		&$-0.36\pm0.11$ (3)\\

\ion{Ba}{2}				&None		&$0.17\pm0.28$ (4)		&$0.25\pm0.09$ (4)		&$0.02\pm0.24$ (5)		&$0.04\pm0.09$ (3)		&$0.11\pm0.14$ (4)\\

\ion{La}{2}				&None		&$0.63\pm0.15$ (2)		&$0.52\pm0.17$ (1)		&$0.56\pm0.19$ (1)		&...					&...				\\

\ion{Nd}{2}				&None		&$0.16\pm0.21$ (2)		&$0.11\pm0.17$ (1)		&$0.09\pm0.14$ (2)		&$0.21\pm0.16$ (1)		&$0.18\pm0.12$ (2)\\

\ion{Eu}{2}				&None		&...					&$0.74\pm0.17$ (1)		&$0.71\pm0.19$ (1)		&...					&$0.70\pm0.17$ (1)				\\

\hline
\label{table:Ab3}
\end{tabular}
\end{minipage}
\end{table*}

\subsubsection{Iron}

The \ion{Fe}{1} and \ion{Fe}{2} abundance derived from the optical data are noticeably different in all stars, with differences ranging from 0.18 to 0.33 dex. It is well known that this is caused by NLTE effects on the \ion{Fe}{1} lines, e.g., \cite{berg2012,lind2012} find that metal-poor RGB stars have \ion{Fe}{1} abundances that are underestimated due to the overionization of iron by the radiation field (effects on \ion{Fe}{2} are negligible). 
NLTE corrections\footnote{Taken from the online database at http://inspect-stars.net/} for NGC 5466-9951 (the star with the largest \ion{Fe}{1} - \ion{Fe}{2} discrepancy) are typically +0.1 dex for a sample of 24 lines.   Assuming this is representative of our samples ($>$100 \ion{Fe}{1} lines per star in the optical), then we apply a global correction to our results (see Table 7).    All elements from all ionization states are computed with respect to \ion{Fe}{1}.

The iron abundances of the individual stars in both clusters agree with the cluster metallicities (see Table 3).
We note the iron abundance spread in NGC 5466 is larger than the 1 $\sigma$ measurement uncertainties, 
with one star at [Fe/H]~=~-2.19 $\pm$ 0.05 dex. \textcolor{black}{We note if the global systematic error of [Fe/H] is considered than it can be seen to be larger than this spread.} This star will be discussed further in Section \ref{sec:origin}.

The iron abundances of the individual stars in both clusters generally agree with the cluster metallicities quoted in the literature. NGC 5024 has an observed global metallicity of -2.10 \citep{b7} and the average found in this work (after NLTE correction) is -2.06~$\pm$~0.13, which we consider in good agreement. NGC 5466 has an observed metallicity of -1.98 \citep{b7} and the average abundance of this work (after NLTE correction) is -1.96~$\pm$~0.13, also in good agreement.

\subsubsection{Carbon and Nitrogen}
\label{sec:carbon}

\begin{figure}
   \centering
    \includegraphics[clip=true,trim = 0 0 0 0,width=0.48\textwidth]{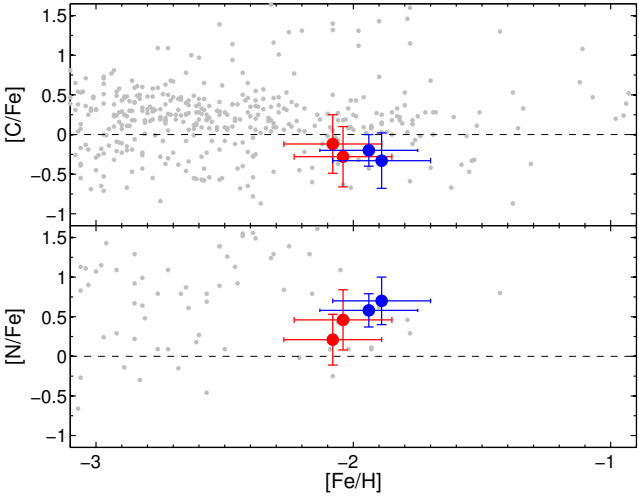}
   \caption{C and N abundances of stars in NGC 5024 and NGC 5466 plotted as a function of \ion{Fe}{1}, compared with Galactic stars from \protect\cite{b28} and \protect\cite{b44} (grey points). Red and blue circles represent NGC 5024 and NGC 5466 stars, respectively.}
   \label{fig:alphaCN}
\end{figure}

Carbon abundances are determined by fitting synthetic spectra to CO lines in the IR (formed by vibration-rotation lines throughout the spectral range 15570 - 16200\AA\hspace{0.20 mm}) and to the atomic \ion{C}{1} line at 16890\AA. Since the C abundances are relatively low, the molecular lines suffer greatly from noise contamination as their line strengths are small. Thus, large uncertainties are shown in the final abundance. The C abundances of this work (derived from 4 of the 5 stars as there was no IR data for NGC 5466-10186) agree with both the ASPCAP results and \cite{b21}; see Table 7. 
The C abundances are compared with a Galactic sample in Fig. \ref{fig:alphaCN} and \textcolor{black}{are} similar to other Galactic stars at the same metallicity.

Nitrogen abundances can be found by fitting synthetic spectra to the CN electronic transition lines located through the IR spectral range 15200 - 15600\AA. The abundances must be determined \textit{after} the C and O abundances are set (refer to section \ref{sec:iranal}) because of the C dependance of the molecular lines. The derived N abundances agree with the ASPCAP values (Table 7) and are consistent with Galactic stars of similar metallicity (see Fig. \ref{fig:alphaCN}). Similar to C, the somewhat large uncertainty is due to relatively weak CN lines suffering from noise contamination.

\subsubsection{$\alpha$-elements}
\label{sec:alpha}
\begin{figure}
   \centering
    \includegraphics[clip=true,trim = 0 0 0 0,width=0.48\textwidth]{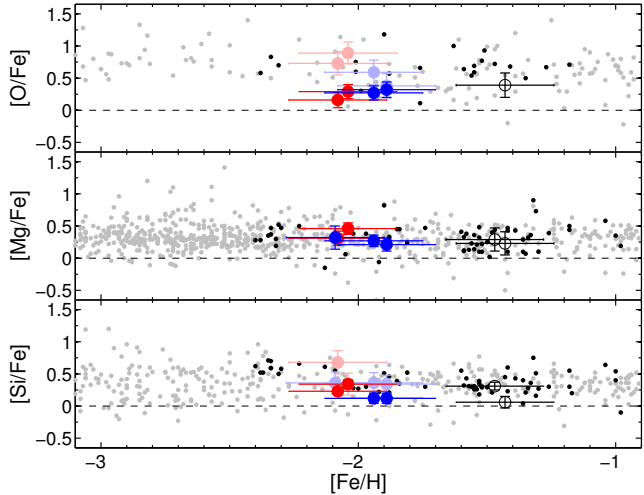}
   \caption{O, Mg, and Si alpha element abundances of stars in NGC 5024/5466 plotted as a function of \ion{Fe}{1}, compared with Galactic stars from the literature. Red circles represent NGC 5024 stars while blue circles are those of NGC 5466 - O and Si abundances come from IR measurements while Mg abundances come from a weighted average between optical and IR (see text). Light gray points represent Galactic distributions of field stars summarized by \protect\cite{b27}, \protect\cite{b28}, and \protect\cite{b44}. Black points represent Galactic GCs, assembled by \protect\cite{b45}. The hollow black points are abundances derived from the standard stars in M3 and M13. The transparent points in the O and Si plots are abundances derived in the optical where only 1 line was available for abundance determination (one transparent point is hidden behind its infrared data point in the Si plot and there is one additional transparent Si point as there is no IR data for that star).}
   \label{fig:alphaA}
\end{figure}

\begin{figure}
   \centering
    \includegraphics[clip=true,trim = 0 0 0 0,width=0.48\textwidth]{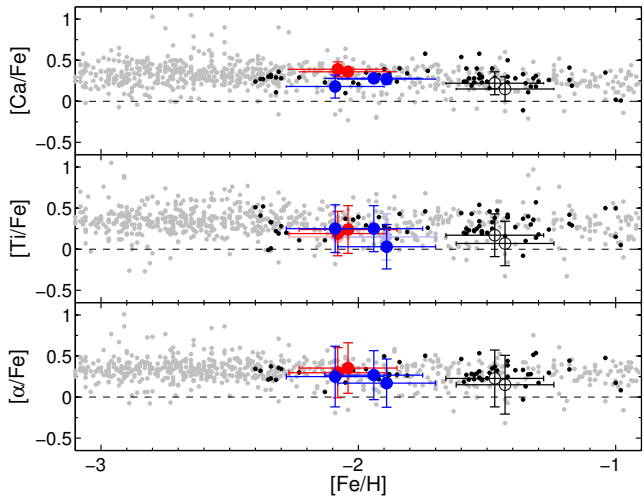}
   \caption{Ca, Ti, and $\alpha$ abundances of stars in NGC 5024/5466 plotted as a function of \ion{Fe}{1}, compared with Galactic stars from the literature as described in Fig. \ref{fig:alphaA}. Only the optical Ti abundance is included, and it is computed as a weighted average between TiI and TiII.}
   \label{fig:alphaB}
\end{figure}

Alpha elements (O, Mg, Si, and Ca) are constructed via the capture of $\alpha$ particles ($^4$He nuclei) during the hydrostatic burning phase of massive stars and $\alpha$-rich explosive nucleosynthesis during Type II supernovae (SN II). Ti is also included as an alpha element as it resembles the other alpha elements in the distribution of [$\alpha$/Fe] ratios in Galactic metal-poor field stars. Comparing the alpha elements to Fe (which dominates the yields from Type Ia SN) then the [$\alpha$/Fe] ratios can inform us on the relative contributions from each type of supernova to the interstellar medium.

\textit{Oxygen}: O abundances are derived from both the optical and the IR. The only line measured in the optical is the forbidden line (6300 \AA\hspace{0.20 mm}, generally quite weak, spanning 13~$<$~EW ~$<$~32~\AA) while O derived from the IR comes from $\sim$12 OH vibration-rotation lines spanning the spectral range 15260 - 16710 \AA\hspace{0.20 mm}. The O abundance derived in 5466-10186 is an upper limit determined from spectrum synthesis. The derived optical and IR O abundances are discrepant (see Table 7), which is not uncommon (e.g., see \citealt{b93,b59}). 
The discrepancy is partially due to a \ion{Ni}{1} blend with the forbidden \ion{O}{1} line in the optical and that there are a small number of optical \ion{O}{1} lines compared to the numerous IR OH lines. 
Also, large systematic effects are found for the OH lines due to temperature uncertainties in model atmospheres \citep{b59,b60}, and 
3D and NLTE effects are not expected to have a major effect on the optical forbidden line \citep{b71,b59}, but may play a role in the line formation of the infrared OH lines.  
\cite{b58} have reproduced the optical abundances from IR spectra of several nearby field giants using 1D and LTE atmospheres in the same spectral windows and line lists used here. 
In this paper, the IR measurements are adopted for oxygen, as shown in Fig. \ref{fig:alphaA}.

\textit{Magnesium}: Mg abundances are derived from both the optical and IR spectra. The optical abundances are mostly determined from three MgI lines at 4703, 5528, and 5711 \AA; the first two lines are generally strong (100~$<$~EW~$<$~140~m\AA) and the third is generally weak (EW$\sim$40 m\AA) while the IR abundances are determined from the three strong lines (EW~$>$~140~m\AA) at 15740, 15748, and 15765 \AA. A weighted average between the optical and IR abundances is computed and shown in Fig. \ref{fig:alphaA}: the resulting abundances agree with most of the other Galactic stars at the same metallicity.  

\textit{Silicon}: Only a single weak Si line (EW~$\sim$~20 m\AA) is available in the optical (at 6155 \AA) and is not in very good agreement with the IR Si abundance(see Table 8). Therefore the final reported Si abundance is only from the $\sim$ 11 lines found in the IR (spanning 15370-16380 \AA, where the lines have varied strength). Fig. \ref{fig:alphaA} shows that the four target stars (those with IR spectra) have Si abundances which are consistent with other Galactic stars at the same metallicity.

\textit{Calcium}: Ca abundances are determined from both optical and IR lines. Although there are significantly more available lines in the optical ($\sim$15, found over the spectral range 5588-6717 \AA, of varied strength) than compared with the IR (2 relatively weak features at 16150 and 16196 \AA) the abundances are found to be in good agreement (see Table 8). The weighted average abundance between the two wavelength regions (computed when both data sets are available) shows Ca in the target stars is comparable to other stars in the Galactic sample at roughly the same metallicity (see Fig. \ref{fig:alphaB}).

\label{sec:tidiff}
\textit{Titanium}: There are numerous TiI and TiII lines available in the optical, spanning the entire wavelength range and of varied line strength; conversely there are very few lines in the IR. Final Ti abundances are determined from the average between TiI and TiII  lines (30-40) in the optical while the single TiI line in the IR (at 15335 \AA) is omitted as it may be less reliable due to line blending with \ion{Fe}{1}. The errors from the abundances of both ionization states are added in quadrature and reported here. Table 8 indicates the IR Ti abundances are in excellent agreement with the optical. Fig. \ref{fig:alphaB} reveals the Ti in the target stars behaves in a similar manner to stars of similar metallicity in the Milky Way.
The abundance discrepancy found between \ion{Ti}{1} and \ion{Ti}{2} (typically $\sim$ 0.2 dex) is a feature which is seen in other work \citep{b68,b69}, but not found here. 
This discrepancy could be explained by NLTE effects associated with \ion{Ti}{1}, described by \cite{b70}\textcolor{black}{. However}
the stellar parameters were outside the range of the target stars in this work and so NLTE corrections are not applied.

\textit{Alpha}: In this work, an $\alpha$-element abundance is defined as the average of Ca, Mg, and Ti,
and is calculated for both the target stars and Galactic distributions taken from the literature. The results are shown in Fig.~\ref{fig:alphaB}. 
Results for the target stars are consistent with the Galactic distribution. Ca and Mg abundances are derived from both optical and infrared data, 
and their weighted averages are shown here. Ti is reported here as the average between \ion{Ti}{1} and \ion{Ti}{2}, and their respective errors are added in quadrature.

\begin{figure}
   \centering
    \includegraphics[clip=true,trim = 0 0 0 0,width=0.48\textwidth]{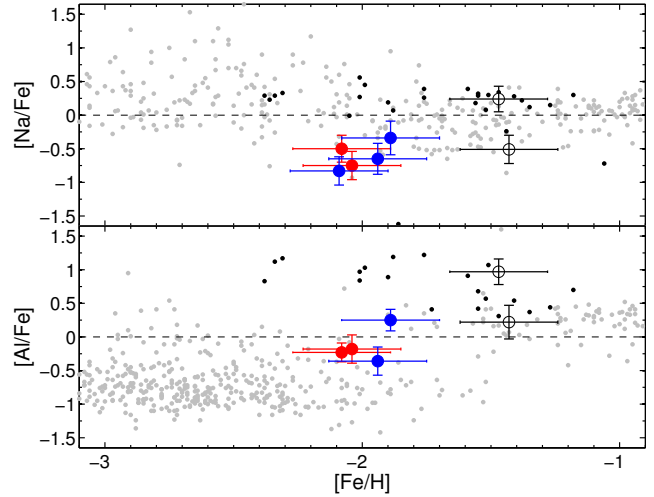}
   \caption{Na and Al abundances of stars in NGC 5024/5466 plotted as a function of \ion{Fe}{1}, compared with Galactic stars from the literature as described in Fig. \ref{fig:alphaA}. \textcolor{black}{The Na abundances in the literature are reported with or without NLTE corrections, depending on the source.} Na abundances for our target stars were determined from optical data (with NLTE corrections) while the Al abundances are from infrared data (explaining why there are only 4 data points).}
   \label{fig:alphaC}
\end{figure}

\textit{Sodium and Aluminum}: Sodium abundances are derived in the optical from the D1 and D2 lines (if the D lines are the only available measurements then lines above 200 m\AA\hspace{0.20 mm} are kept) and the line at 5688 \AA\hspace{0.20 mm} (which is generally weak, $\sim$~20~m\AA). Sodium is known to have a range of NLTE effects that are line dependent, but also depend on the Fe and $\mathrm{T}_{\mathrm{eff}}$ of the models. Corrections have been applied following \cite{b72}\footnote{Adopted from the online database at http://inspect-stars.net/}, where NLTE corrections ranged from -0.05 dex (for the 5688 \AA\hspace{0.20 mm} line) up to -0.29 for the D lines. Abundance measurements are shown in Fig. \ref{fig:alphaC}, almost all of which are relatively low compared to the Galactic GCs. 
The GC abundance average would include the second-generation Na-enhanced stars.

Aluminum lines were only found in the higher metallicity standard stars at optical wavelengths. Conversely, there are three strong (EW~$>$~100m\AA) \ion{Al}{1} lines in the IR at 16718, 16750, and 16760 \AA. The Galactic distribution of Al in stars around this metallicity is quite dispersed and the NGC 5466/5024 stars fall more or less within the Galactic field star distribution.

\subsubsection{Iron-peak Elements}

\begin{figure}
   \centering
    \includegraphics[clip=true,trim = 0 0 0 0,width=0.48\textwidth]{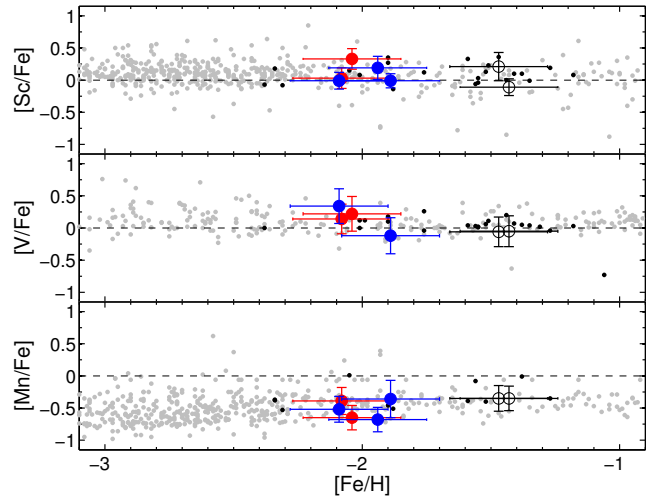}
   \caption{Sc, V, and Mn abundances of stars in NGC 5024/5466 plotted as a function of \ion{Fe}{1}, compared with Galactic stars from the literature as described in Fig. \ref{fig:alphaA}.}
   \label{fig:alphaD}
\end{figure}

\textit{Scandium}, \textit{vanadium}, and \textit{manganese} are derived exclusively from optical data, using 1-6 spectral lines (per element) and are shown in Fig. \ref{fig:alphaD}.  Scandium is found entirely from singly ionized lines and two of the five target stars have slightly high Sc abundances. As noted in Section \ref{sec:hfs}, HFS corrections were negligible for Sc. V was derived for two stars from each target cluster from generally weak lines (EW~$\sim$~20 m\AA). \textcolor{black}{Hyperfine} structure corrections for V were calculated and applied to each individual star (typically -0.07 dex, following the references discussed in Section \ref{sec:hfs}). The V abundance appears consistent with the Galactic sample although there is a significant scatter in V for the NGC 5466 stars; these NGC 5466 abundances are measured from a single V line in each star and the scatter agrees within the errors therefore this scatter may not be real. The Mn lines varied in strength and were found to have significant hyperfine structure corrections (ranging from -0.06 to -0.21 dex, depending on the star); after applying these corrections (again following the references in Section \ref{sec:hfs}) the Mn in the target stars appears to follow the Galactic distribution at their respective metallicities. \cite{b84} show that a metal-poor star ([Fe/H]$\sim$-2.5) with similar stellar parameters to the targets in this work and has a \textit{total} NLTE correction of +0.44 dex, however there are no stars in their sample that share the same parameter space as our standard stars. We do not apply this correction because this would not allow a comparison with our standard stars and it is unclear whether the correction of \cite{b84} applies to the same lines of this work. However, it should be noted that NLTE effects are strong for Mn lines within the stellar parameters of our sample. 

\begin{figure}
   \centering
    \includegraphics[clip=true,trim = 0 0 0 0,width=0.48\textwidth]{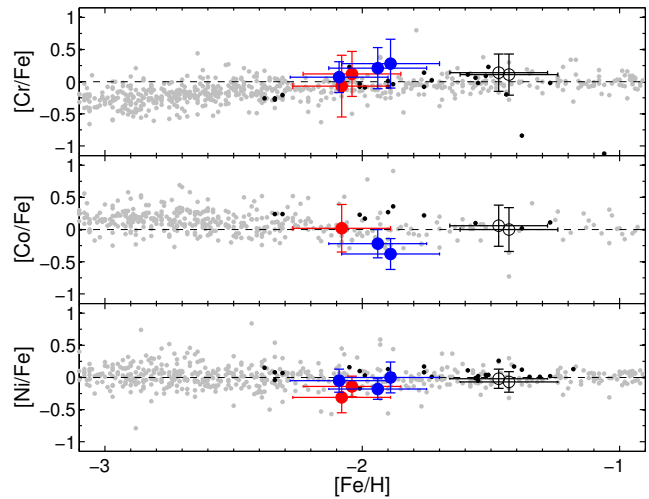}
   \caption{Cr, Co, and Ni abundances of stars in NGC 5024/5466 plotted as a function of \ion{Fe}{1}, compared with Galactic stars from the literature as described in Fig. \ref{fig:alphaA}. The Cr abundance is reported as a weighted average between \ion{Cr}{1} and \ion{Cr}{2}, where NLTE corrections have been applied to each ionization state.}
   \label{fig:alphaE}
\end{figure}

\textit{Chromium}, \textit{cobalt}, and \textit{nickel} are also derived exclusively in the optical. The Cr abundance is reported as a line weighted average between 6-10 \ion{Cr}{1} and \ion{Cr}{2} lines found over the spectral range 4550-6330 \AA\hspace{0.20 mm} and of varying line strength.  The discrepancy between \ion{Cr}{1} and \ion{Cr}{2} is apparent, which can be attributed to NLTE effects of \ion{Cr}{1}, e.g. \cite{b62}. NLTE corrections were found by locating stars of a similar metallicity to this sample, looking at common chromium lines between the two datasets, and then extracting the NLTE abundance correction. The NLTE correction was applied line by line to the abundances in this sample, with corrections typically found to be 0.22 dex for target stars and 0.10~dex for the standard stars; the \ion{Cr}{2} correction also applied in the same manner and was generally found to be much smaller ($\sim$ 0.05 dex). The Cr abundance in the NGC 5024/5466 stars is slightly higher than the Galactic distribution (see Fig. \ref{fig:alphaE}), however given the NLTE corrections and 1-$sigma$ uncertainties then we consider our results in good agreement with the other Galactic globular cluster data.
Co is found in both cluster stars from weak lines (EW~$\sim$~20 m\AA). In NGC 5024 Co is found in one of the stars, specifically from 2 lines at 5483 and 5647 \AA\hspace{0.20 mm} and agrees with the Galactic distribution (see Fig. \ref{fig:alphaE}). In both the NGC 5466 stars Co is determined from 1 line at 5483 \AA. \cite{b92} find Co NLTE corrections for the same star used to find Mn NLTE corrections in \cite{b84}. They find Co has high NLTE corrections at low metallicities, similar to Mn, and find corrections of 0.64 dex for the star that shares the same parameter space with this work. Co NLTE corrections are not applied to this work for the same reasons discussed with Mn. Finally, Ni abundances are derived from 5-8 optical lines of varied strength and are in good agreement with other Galactic stars at this metallicity.

\subsubsection{Copper and Zinc}

\begin{figure}
   \centering
    \includegraphics[clip=true,trim = 0 0 0 0,width=0.48\textwidth]{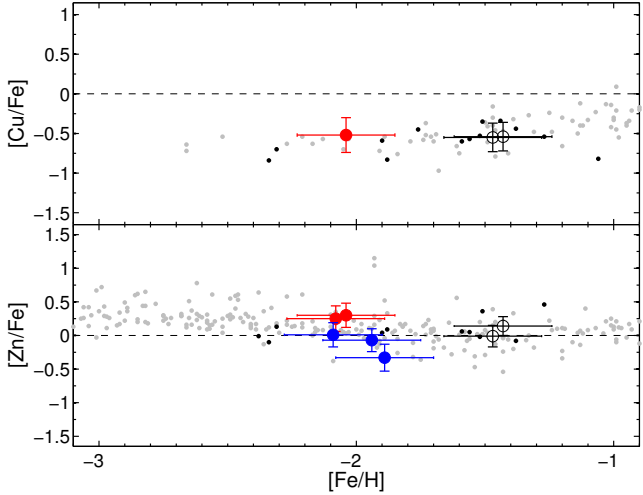}
   \caption{Cu and Zn abundances of stars in NGC 5024/5466 plotted as a function of \ion{Fe}{1}, compared with Galactic stars from the literature as described in Fig. \ref{fig:alphaA}. Also included are Cu and Zn abundances from \protect\cite{b29}, also plotted as light gray points. The Cu line used to compute the Cu abundance is quite weak and only detectable in NGC 5024-50371.}
   \label{fig:alphaF}
\end{figure}

\textit{Copper} and \textit{zinc} measurements are relatively scarce in the literature, thus an additional source is added \citep{b29} for comparison. Copper is only detected in one of the NGC 5024 stars (from the single weak line at 5105 \AA) and not in NGC 5466. Zinc is measured in all the target stars from the neutral line at 4722 \AA (typically with an EW of 40 m\AA); the NGC 5466 stars show a trend slightly lower (see Fig. \ref{fig:alphaF}). The NGC 5466 Zn abundance was noted to be significantly low in a detailed abundance analysis of a single anomalous cepheid star reported by \cite{b30}. They argue that since the low [Zn/Fe] ratio should reflect the primordial abundance of their target star and predicted other stars would also be low in Zn.

\subsubsection{Neutron-capture Elements}

The neutron-capture elements come from one of two processes: either the slow process (or s-process) via AGB stellar winds or through the rapid process (r-process) during SNe II nucleosynthesis. \cite{burris2000} have shown that in the Sun 97 $\%$ of the Eu abundance originates from the r-process, whereas at least 72\% of the Y, Ba, and La in the Sun come from the s-process. Thus these elements are excellent indicators as to which of these two types of processes have been involved within the environment of the cluster.

\begin{figure}
   \centering
    \includegraphics[clip=true,trim = 0 0 0 0,width=0.48\textwidth]{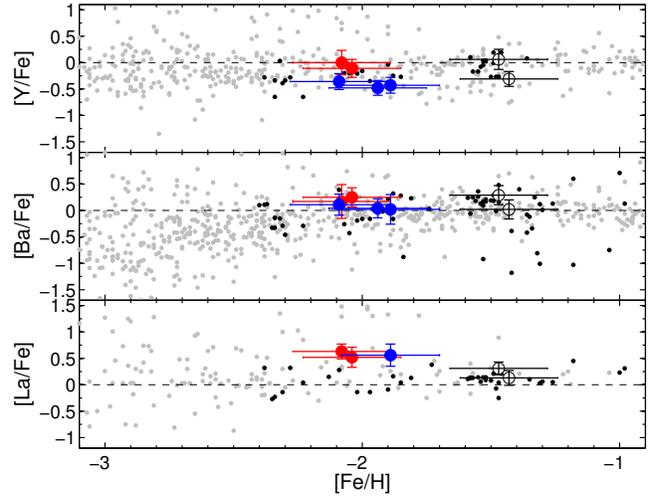}
   \caption{Y, Ba, and La abundances of stars in NGC 5024/5466 plotted as a function of \ion{Fe}{1}, compared with Galactic stars from the literature as described in Fig. \ref{fig:alphaA}.}
   \label{fig:alphaG}
\end{figure}

\textit{Yttrium} is detected in all five target stars from 3-4 singly ionized lines of varied strength in the optical data. Y in both clusters is consistent with the Galactic comparison sample. The abundance in the two NGC 5024 stars is higher than those in NGC 5466 (see Fig. \ref{fig:alphaG}). The Y can be seen to have little star to star variations within each cluster and is also in agreement with the standard stars in M3 and M13.

\textit{Barium} is detected in all five target stars from 3-5 strong (EW~$>$~100 m\AA) singly ionized lines in the optical. Fig. \ref{fig:alphaG} shows the small star to star dispersion of Ba within each cluster and it also shows the Ba is in good agreement with the M3/M13 standard stars.

\textit{Lanthanum} is detected from only 1-2 weak, singly ionized optical lines in the NGC 5024 stars and one NGC 5466 star. 
These La abundances appear to be slightly higher than our M3/M13 stars and the Galactic distribution (see Fig. \ref{fig:alphaG}).

\begin{figure}
   \centering
    \includegraphics[clip=true,trim = 0 0 0 0,width=0.48\textwidth]{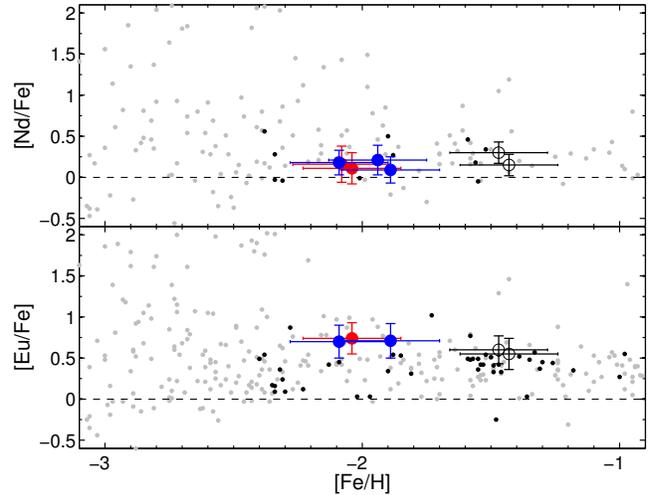}
   \caption{Nd and Eu abundances of stars in NGC 5024/5466 plotted as a function of \ion{Fe}{1}, compared with Galactic stars from the literature as described in Fig. \ref{fig:alphaA}.}
   \label{fig:alphaH}
\end{figure}

\textit{Neodymium} abundances are measured from 1-2 weak lines from the optical data (at 5249 and 5319 \AA) for all five of the target stars. Nd measurements show very little star to star variation within each cluster and also agree with our standard stars (see Fig. \ref{fig:alphaH}).

\textit{Europium} is only detected in three stars (two in NGC 5466 and one in NGC 5024) from the weak 6645 \AA\hspace{0.20 mm} line in the optical (EW~$\sim$~15 m\AA for all three stars). Fig. \ref{fig:alphaH} shows the measured Eu abundance in both clusters is relatively higher than the Galactic distribution.


\begin{figure}
   \centering
    \includegraphics[clip=true,trim = 5 0 0 0,width=0.48\textwidth]{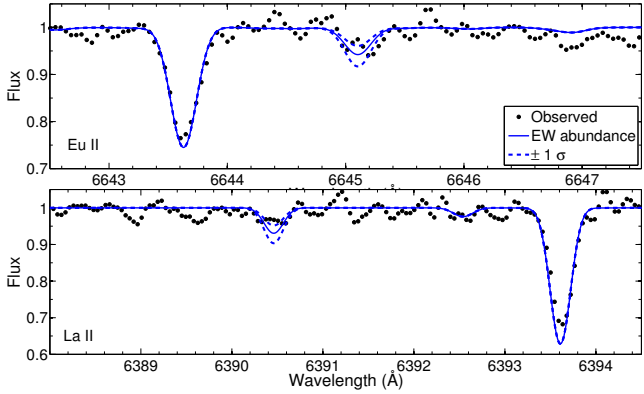}
   \caption{Top: synthetic abundance fit to the single Eu line in NGC 5466-9951; the 1 sigma errors here are 0.19 dex. The spectrum synthesis agrees with the EW analysis abundance. Bottom: synthetic fit to the weak La line in the same star, along with 1 sigma errors (0.19 dex); the synthetic abundance agrees with the EW abundance.}
   \label{fig:eu}
\end{figure}

To validate the accuracy of the Eu abundance derived from the EW analysis we synthesize the weak line at 6645 \AA\hspace{0.20 mm} for a single star (NGC 5466-9951) and compare. Fig. \ref{fig:eu} shows a synthetic spectrum for the EW Eu abundance with 1 sigma errors. Also shown is the synthetic fit to the weak La line in the same star, where the synthetic spectrum is created with the EW La abundance (also with 1 sigma errors). 


\section{Discussion}

\subsection{Infrared Abundance Comparison with Optical and Literature Abundances}
\label{sec:litcomp}

\begin{figure*}
   \centering
    \includegraphics[clip=true,trim = 0 0 0 0,width=1\textwidth]{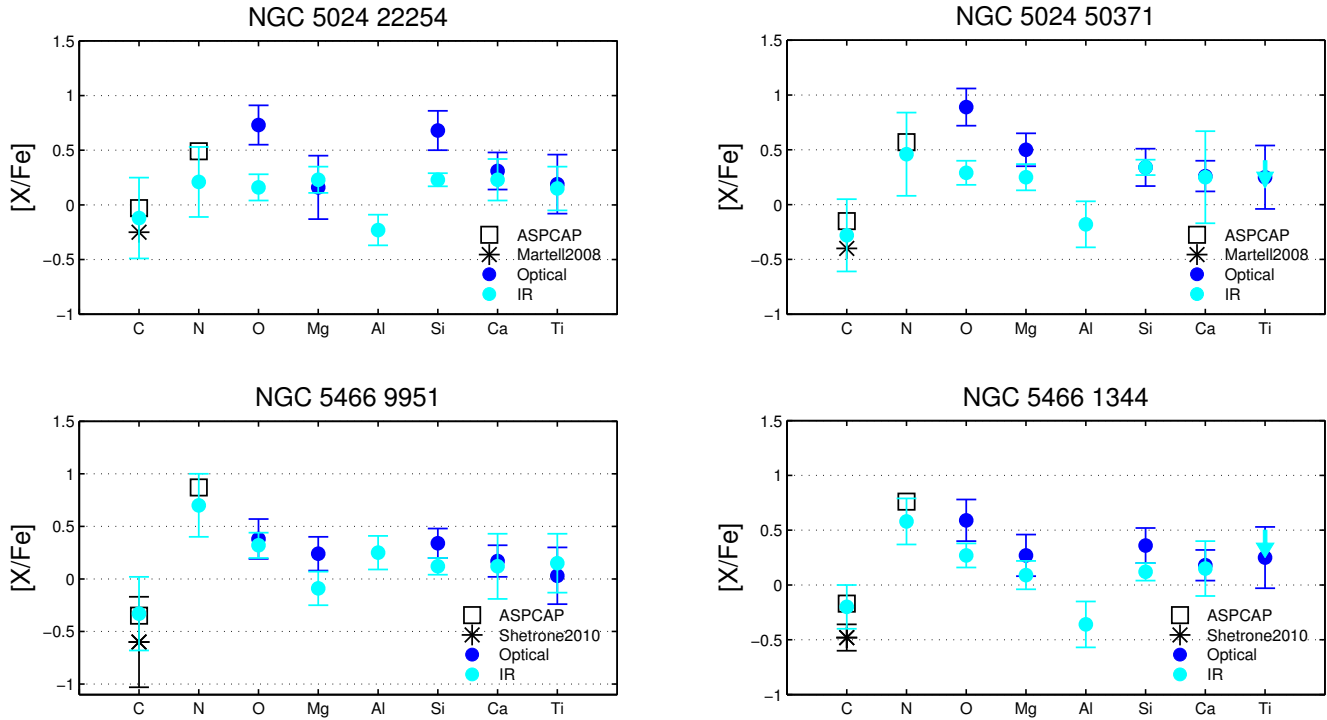}
   \caption{Abundance comparison of overlapping elements between target stars. The blue points are from the optical observations of this work, the cyan points from infrared APOGEE observations, the square points from APOGEE's abundance pipeline ASPCAP, and the asterisk points from \protect\cite{b46} (NGC 5024), \protect\cite{b21} (NGC 5466). \textcolor{black}{The errors reported here reflect both random and systematic uncertainties.} The ASPCAP points and the points of \protect\cite{b46} are plotted without errors, as the reported values in each case are negligibly small and perhaps do not reflect the true spread in the measurements. In general, the results are consistent except for O and Si in NGC 5024-22254 and for O in NGC 5024-50371.}
   \label{fig:Ab_comp}
\end{figure*}

There are 7 elements from the sample of stars in this work (including standard stars) which have abundances derived from both the optical and the IR: Fe, O, Mg, Al, Si, Ca, and Ti \textcolor{black}{(although Al is only derived in the optical for our standard stars, it is included here to demonstrate it is among the list of elements that \textit{can} be compared between the two wavelength analyses)}. Tables \ref{table:Ab1} and \ref{table:Ab2} summarize the reported abundances and show the excellent agreement of Fe derived from the optical and the IR. \textcolor{black}{The uncertainties shown here reflect both random and systematic errors.} The ASPCAP (see Section \ref{sec:aspcapref}) Fe abundances are also in excellent agreement, even though the model atmospheres used to compute the abundances in this work have different parameters (although within the errors, see Table \ref{table:inputParam}).
In Fig. \ref{fig:Ab_comp}, we compare the optical and IR abundance results for the remaining six elements that are found in both wavelength regions 
(Fe is excluded as the abundances are shown as [X/Fe]), as well as C and N. 
There are reported abundances for C and N from ASPCAP and for C from \cite{b21} and \cite{b46}.
The IR C abundance in NGC 5024 is in excellent agreement with ASPCAP and \cite{b46}. Similarly the IR C abundance in NGC 5466 is consistent both with ASPCAP and \cite{b21}. The N abundance in both clusters also shows consistency with ASPCAP. The O abundance derived in the IR is much lower than that of the optical for the two stars in NGC 5024; this discrepancy is discussed in section \ref{sec:alpha} and we favour the IR results. The rest of the element abundances are consistent, with the exception of Si in one star;  we favour the IR abundances for Si since there are only 1-2 measured 
lines in the optical but many more lines in the IR.

\subsection{r and s-process Contributions}

\begin{figure}
   \centering
    \includegraphics[clip=true,trim = 0 0 0 0,width=0.48\textwidth]{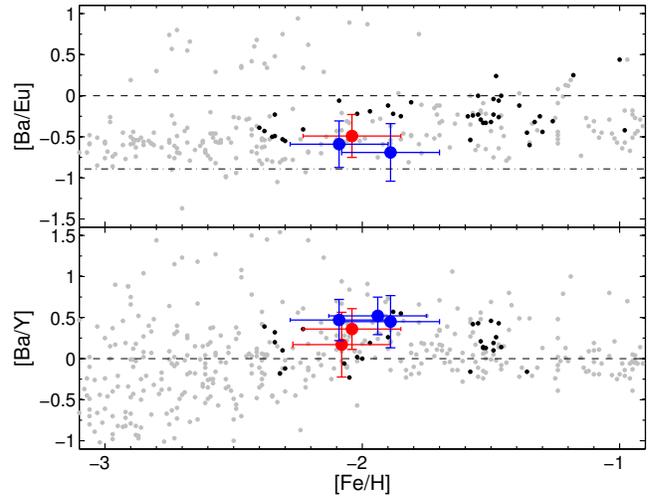}
   \caption{[Ba/Eu] vs [Fe/H] and [Ba/Y] for the NGC 5024/5466 stars with available abundances. The lower dashed-dotted line in the top plot represents the lowest [Ba/Eu] ratio possible, where only the r-process contributes to these elements \protect\citep{burris2000}.}
   \label{fig:BaEuBaY}
\end{figure}

Europium is a used as proxy for the r-process (based on an analysis of the solar system distribution of the heavy elements, e.g., \citealt{arlandini99,burris2000}), whereas Y, La, and Ba are formed in a number of nucleosynthetic sites ranging from core collapse SN to the s-process during the thermal pulsing in AGB stars.  \cite{gallino1990} and \cite{busso1999} showed that the s-process yields in AGB stars are metallicity dependent, such that low metallicity AGB stars will bypass the first s-process peak elements in favour of the second and third peak elements due to a lack of iron seed nuclei.     In Galactic field stars, the increase in these elements from the s-process can be seen between metallicities of $-3 <$ [Fe/H] $< -2$ (e.g., \citealt{mcwill95,b65}). 
Fig. \ref{fig:BaEuBaY} shows [Ba/Eu] and [Ba/Y] as a function of metallicity in Galactic globular clusters and field stars; 
the r-process yield in [Ba/Eu] from \cite{burris2000} is also shown.  
The slightly elevated [Ba/Eu] values in our two globular clusters are consistent with a mild s-process contribution, and the [Ba/Y] values suggest those contributions are from metal poor AGB stars.

\subsection{Evidence for Mixing}

\begin{figure*}
   \centering
    \includegraphics[clip=true,trim = 0 0 0 0,width=0.45\textwidth]{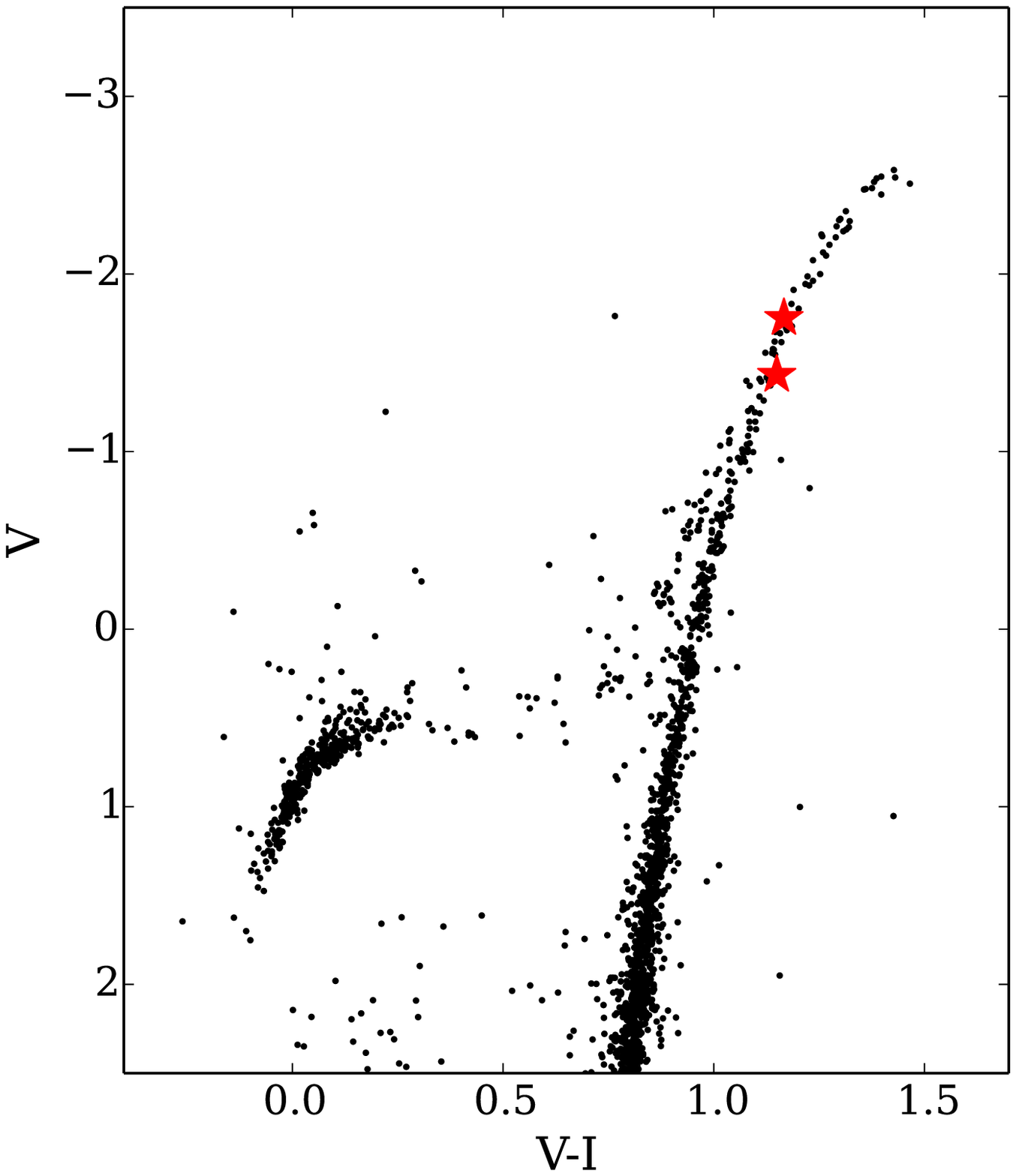}\includegraphics[clip=true,trim = 0 0 0 0,width=0.45\textwidth]{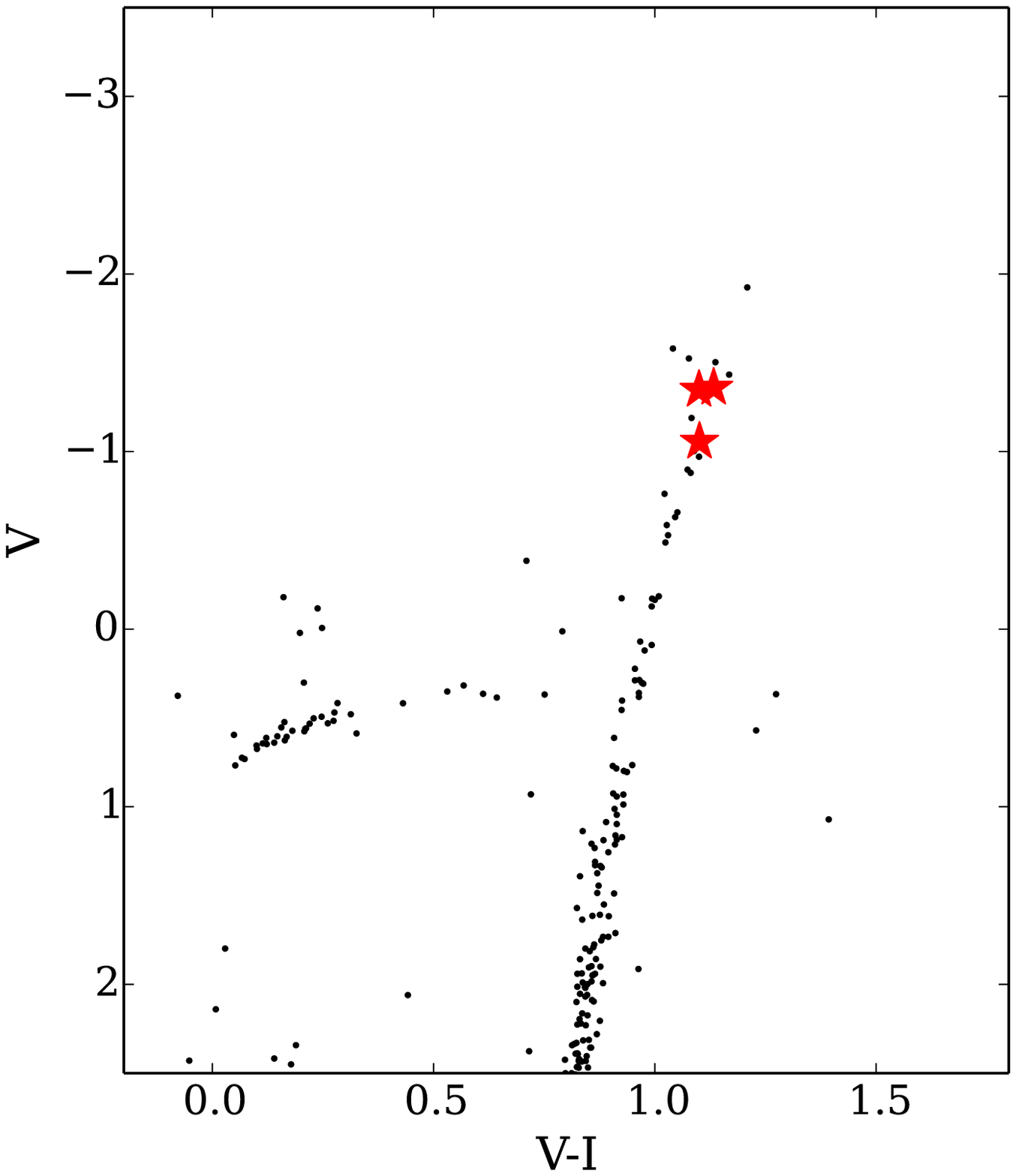}
   \caption{Colour-magnitude diagrams for the globular clusters NGC 5024 (left) and NGC 5466 (right), with photometry taken from the ACS Survey of Galactic Globular Clusters \protect\citep{saraj07,anderson08}. \textcolor{black}{The symbols in red are the target stars from this work.}} 
   \label{fig:cmds}
\end{figure*}

The locations of the target stars on their colour-magnitude diagrams (CMDs) are shown in Fig. \ref{fig:cmds};  photometry is from \cite{saraj07,anderson08}. All of our targets are bright stars located near the tip of the red giant branch, above the RGB bump, where mixing of CNO-cycled H-burning gas is expected (e.g., \citealt{suntz, smithmartell, smithbriley}, \citealt{Charbonnel1998}, \citealt{sweig, charbonnel95,bellman2001,deniss2003,carretta2005}). This will have the effect of elevated N (by $\sim$ 0.5 dex), slightly depleted C (by $\sim$0.2 dex), and even slighter depletions of oxygen ($< 0.1$ dex) from their initial abundances \citep{carretta2005}. Assuming the globular clusters started with solar C and N, then the [(C+N)/Fe]  abundances will remain $\sim$solar (we ignore the very small change in oxygen that is expected from its enhanced, Galactic plateau, initial value).  The [(C+N)/Fe] values are shown in Fig. 19, and are consistent with standard mixing as the sums remain near solar.  NGC5466-9951 also shows slight enhancements in Na and Al suggestive of primordial variations or pollution within the cluster (e.g., \citealt{b43,carretta2009}).  This suggests that NGC 5466-9951 may be a member of a second generation population in this cluster.

\begin{figure}
   \centering
    \includegraphics[clip=true,trim = 0 0 0 0,width=0.45\textwidth]{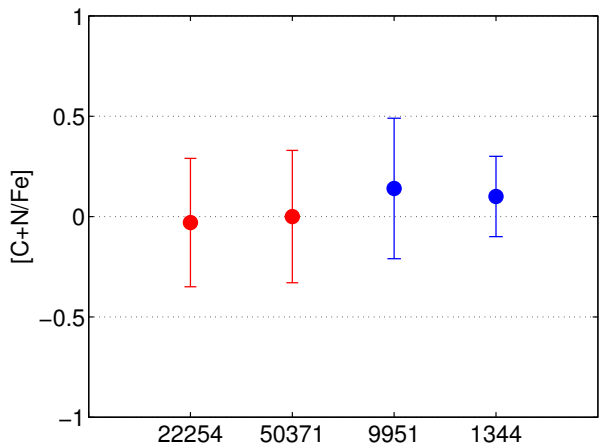}
   \caption{[C+N/Fe] abundances to the NGC 5024/5466 stars where the data was available. The red and blue points are those of NGC 5024 and NGC 5466, respectively.}
   \label{fig:CNO}
\end{figure}

\subsection{NGC 5024/5466 /origins}
\label{sec:origin}
NGC 5024 and NGC 5466 do not show any distinct abundance differences with GC and field stars within the MW, although at these low metallicities it is difficult to discern whether the abundances of these clusters would show extra-galactic signatures. As previously described, this is because at low metallicities GCs like M54 do not show chemical abundance patterns that stand out with GCs in the MW, even though it has physical associations with a dwarf remnant. However, M54 and $\omega$ Cen \textit{do} show an Fe metallicity spread (i.e. 0.19 dex rms scatter for M54), and are among the few GCs known to show such a spread \cite{b83}; where it is also argued this spread could be caused by large and repeated bursts of star formation. Interestingly, NGC 5466 also shows a spread in Fe abundance found from the three stars in this work (0.09 dex rms scatter \textcolor{black}{from the random errors; if one considers systematics then the scatter is smaller than this value}). The velocity dispersion is relatively large in this sample (~12 km/s), which is also seen to be large by \cite{b21} (17 km/s, although from low resolution spectra). It is possible that the one star NGC 5466-10186 that causes most of this velocity dispersion is a non-member, however the position of this star on the CMD and lack of foreground contamination towards this cluster makes this unlikely.  It is also possible that NGC 5466-10186 is in a binary system which may have affected its apparent radial velocity and could also explain the slightly lower metallicity we have found if the binary companion has artificially increased its continuum flux levels.   Before interpreting the velocity and Fe dispersions in NGC5466 in terms of its origins or physical mass, more stars should be examined in this cluster.

\section{Summary and Conclusions}

A detailed chemical abundance analysis has been performed for two RGB stars in NGC 5024 and three RGB stars in NGC 5466.   In this analysis, we have found:

\begin{enumerate}
  \item The optical and infrared abundances are in good agreement for the elements Fe, Mg, Si, Ca, and Ti. There is a discrepancy in the oxygen abundances, which has been seen before in the literature.   We favour our IR oxygen abundances, determined from several OH features.
  \item The neutron-capture elements are mildly enriched in s-process yields from metal-poor AGB stars.
  \item The stars in both clusters exhibit CNO mixing, as evidenced from the enhanced N, depleted C, and solar [(C+N)/Fe] ratios.   One star NGC5466-9951 shows slight enhancements in Na and Al as well, which suggests it may be a member of a second generation population.  
  \item Both globular clusters show element ratios that are similar to other Galactic clusters at this metallicity, however NGC5466 may have a larger spread in the iron abundances and a larger velocity distribution.   This may be due to one star (NGC5466-10186) in a binary system, or it may represent a physical property of this cluster.   Other clusters with metallicity and velocity dispersions include Omega Cen and M54, both accreted from the Sgr dwarf galaxy. 
\end{enumerate}

We conclude that abundances derived from both the optical and infrared regions complement each other in a stellar atmospheres analysis, and that NGC5024 and NGC5466 appear to be similar to the majority of Galactic globular clusters.  Further spectroscopic analysis of NGC5466 is needed to confirm its apparent metallicity and velocity dispersions.

\vspace{10mm}
We are grateful to the SDSS APOGEE team for making their spectra available in addition to their ASPCAP stellar pipeline results.  MPL and KAV acknowledge funding from an NSERC Discovery Grant, and CMS would like to thank NSERC for a Vanier Graduate Fellowship.

\end{document}